\documentclass[aps,prb,amsmath,amssymb,superscriptaddress,nofootinbib,longbibliography,onecolumn]{revtex4-2}
\usepackage{graphicx}
\usepackage{mathrsfs}
\usepackage{bm}
\usepackage{textcomp,color}
\usepackage{xcolor}
\usepackage[colorlinks=true,urlcolor=blue,citecolor=blue,linkcolor=blue,bookmarks=false, pdfstartview={FitH}]{hyperref}
\usepackage[capitalize]{cleveref} 
\usepackage{amsmath}
\usepackage[normalem]{ulem} 

\newcommand{\beq}{\begin{equation}}
\newcommand{\eeq}{\end{equation}}
\newcommand{\bea}{\begin{eqnarray}}
\newcommand{\eea}{\end{eqnarray}}
\newcommand{\bse}{\begin{subequations}}
\newcommand{\ese}{\end{subequations}}
\newcommand{\nn}{\nonumber}
\newcommand{\bwt}{\begin{widetext}}
\newcommand{\ewt}{\end{widetext}}

\newcommand{\e}{\epsilon}

\newcommand{\bk}{{\bf k}}

\newcommand{\bq}{{\bf q}}
\newcommand{\br}{{\bf r}}
\newcommand{\bv}{{\bf v}}

\newcommand{\bE}{{\bf E}}

\newcommand{\bJ}{{\bf J}}

\begin{document}

\title{Resonant Edelstein and inverse-Edelstein effects, charge-to-spin conversion, and spin pumping from chiral-spin modes}
\author{Mojdeh Saleh}
\affiliation{Department of Physics, Concordia University, Montreal, QC H4B 1R6, Canada}
\author{Abhishek Kumar}
\thanks{Current address: Department of Physics, Indian Institute of Technology Jammu, Jammu 181221, India}
\affiliation{Département de Physique and Institut Quantique, Université de Sherbrooke, Sherbrooke, Québec J1K 2R1, Canada}
\author{Dmitrii L. Maslov}
\affiliation{Department of Physics, University of Florida, Gainesville, Florida, 32611, USA}
\author{Saurabh Maiti}
\affiliation{Department of Physics, Concordia University, Montreal, QC H4B 1R6, Canada}
\affiliation{Centre for Research in Multiscale Modelling, Concordia University, Montreal, QC H4B 1R6, Canada}
\date{\today}
\begin{abstract}
Spin-orbit coupling in systems with broken inversion symmetry gives rise to the Edelstein effect, which is the spin polarization induced by an electric field or current, and the inverse-Edelstein effect (also known as the spin-galvanic effect), which is the electric current induced by an oscillatory magnetic field or spin polarization. At the same time, an interplay between spin-orbit coupling and electron-electron interaction leads to a special type of collective excitations--chiral-spin modes--which are oscillations of spin polarization in the absence of a magnetic field. As a result, both Edelstein and inverse-Edelstein effects exhibit resonances at the frequencies of chiral-spin collective modes. Here, we present a detailed study of the effect of electron correlation on the resonances in Edelstein and inverse-Edelstein effects in a single-valley two-dimensional electron gas and in a multi-valley Dirac system with proximity-induced spin-orbit coupling. While the chiral-spin modes involve both in-plane and out-of-plane oscillations of spins, we show that only the in-plane modes are responsible for the above resonances. In the multi-valley system, electron correlation splits the in-plane modes into two. We study the spectral weight distribution between the two modes over a large parameter space of intra- and inter-valley interactions. Finally, we demonstrate that using the chiral-spin modes one can get a resonant enhancement of charge-to-spin conversion and gain a directional control of the injected spins in the spin-pumping process, both of which are relevant to spintronics.
\end{abstract}

\maketitle
\tableofcontents

\section{Introduction}\label{Sec:Introduction}
In systems with inversion symmetry, electron spin couples to its momentum in two ways. One is Mott skew scattering \cite{Mott1965}, where electrons scatter from heavy-atom impurities with an amplitude that depends on the angle between the momentum and spin. The other one is the Elliot-Yafet mechanism \cite{Elliot1954,YAFET1963}, where Bloch electrons residing in a spin-orbit coupled energy band scatter from (not necessarily heavy) impurities in a spin-sensitive manner. On the other hand, breaking of inversion symmetry lifts the two-fold degeneracy of electron states (while still preserving the Kramers degeneracy), thus enabling direct coupling between electron momenta and spins. Examples of spin-orbit coupling (SOC) in systems with broken inversion symmetry are Dresselhaus \cite{Dresselhaus1955}, Rashba \cite{Rashba_Sheka,Bychkov1984}, and, in multi-valley systems, the valley-Zeeman (VZ) (also known as Ising) effects\cite{Kormanyos2014}. 

Spin-orbit interaction of both types (with and without inversion symmetry) leads to several spin-dependent transport phenomena. These include the spin Hall effect~\cite{DYAKONOV1971a,DYAKONOV1971b,Hirsch1999}, which is the induced spin accumulation transverse to the electric current (observed via Kerr microscopy~\cite{Kato2004a} and electroluminescence~\cite{Wunderlich2005} in GaAs-based heterostructures), and the inverse spin Hall effect ~\cite{Averkiev1983}, which is the electric current induced by a non-uniform spin injection (observed via optical induction of spin polarization ~\cite{Bakun1984,Tkachuk1986}, also in GaAs-based heterostructures). Broken inversion symmetry, in particular, leads to a current-induced spin polarization in systems with odd-in-momentum spin-splitting of the electron states\cite{Aronov:1989}, known in the context of two-dimensional electron gases (2DEGs) as the Edelstein effect~\cite{EDELSTEIN1990}, which has been observed in GaAs heterojunctions~\cite{silov2004} and possibly also in strained InGaAs~\cite{Kato2004}. Systems with broken inversion symmetry also exhibit the inverse-Edelstein/spin-galvanic effect (observed in GaAs heterojunctions~\cite{Ganichev2002} and at Bi/Ag interface~\cite{sanchez2013}, and explained subsequently in Ref.~\cite{shen2014}), where an inhomogeneous spin distribution drives electric current\cite{GANICHEV2024}.

Apart from the steady-state effects, described above, there are noteworthy dynamical effects that arise in systems with broken inversion. One of them is the dynamical spin accumulation at the 2DEG boundaries \cite{Duckheim2009}. Then there is the electric-dipole spin resonance (EDSR) effect \cite{Rashba2003,Efros2006,Duckheim2006}, in which the electric($E$)-field of an electromagnetic wave couples directly to electron spins via spin-orbit interaction. This effect was observed in a variety of systems: bulk semiconductors~\cite{Bell1962,McCombe1967}, quantum wells~\cite{Schulte2005}, quantum dots~\cite{Nowack2007,Pioro-Ladrière2008}, and nanowire-based qubits~\cite{Nadj-Perge2010}. In all previous experiments, spin splitting was induced by a static magnetic($B$)-field. However, it has been predicted that SOC-induced intrinsic splitting of the Bloch states leads to resonances from the so-called chiral-spin modes (CSMs) even in the absence of the static magnetic field~\cite{shekhter2005,ashrafi2012,maiti2015a,maiti2016,kumar2017}. The resonances from these collective modes can be probed by both oscillatory $E$- and $B$-fields. Hitherto not detected via EDSR, resonances due to CSMs have been observed by resonant electronic Raman spectroscopy in zero field in topological surface states of Bi$_2$Se$_3$~\cite{kung2017}, and in finite fields in CdTe quantum wells~\cite{Perez2007,Baboux2013,Baboux2015,Perez2016, Karimi2017}. CSMs have also been predicted to exist both at finite and zero fields in multi-valley systems with proximity-induced SOC, such as graphene on a transition-metal-dichalcogenides (TMD) substrate~\cite{Raines2021,kumar2021}.

Collective modes encode important information about the nature of excitations and correlations in a system. With the rising interest in terahertz (THz) spectroscopies of quantum materials\cite{Bera2021} and in ultrafast spin-based devices \cite{Vaidya2020,Li2020}, it becomes essential to not only know the dispersion relations of collective modes, but also to know how they respond to various experimental probes. In this light, the goal of the present paper is to investigate how CSMs respond to oscillatory $E$ and $B$-fields and study how they modify the SOC induced effects mentioned above. Linear response of a metal to oscillatory $E$ and $B$-fields can be written in terms of the induced magnetization $M_\alpha$ and induced current $J_\alpha$ ($\alpha,\beta\in\{x,y\}$) as
\begin{subequations}
\label{eq:genResp0}
\bea
M_\alpha(\bq,\Omega) &=& \chi^s_{\alpha\beta}(\bq,\Omega)B_\beta(\bq,\Omega) + \sigma^{\rm ME}_{\alpha\beta}(\bq,\Omega)E_\beta(\bq,\Omega),\label{eq:genResp}\\
J_\alpha(\bq,\Omega) &=&\sigma_{\alpha\beta}(\bq,\Omega)E_\beta(\bq,\Omega)+\chi^{\rm JB}_{\alpha\beta}(\bq,\Omega)B_\beta(\bq,\Omega),\label{eq:genRespb}
\eea
\end{subequations}
where $\sigma^{\rm ME}$ and $\chi^{\rm JB}$ are the \textit{cross-response} tensors that arise solely due to SOC, as opposed to the direct-response tensors, spin susceptibility($\chi^s$) and conductivity($\sigma$), which exist already without SOC. We use the convention where the repeated index is summed over. The spin susceptibility $\chi^s$ entails the conventional electron spin resonance (ESR) which is induced by applying an additional static magnetic field, as well as the chiral-spin resonance (CSR) \cite{shekhter2005,maiti2016} which is the zero-field spin resonance due to CSMs. The conductivity $\sigma$ entails both the Drude part and the EDSR effect \cite{maiti2015a,kumar2021}. In this work, we analyze the cross-response terms in Eqs.~\eqref{eq:genResp} and Eqs.~\eqref{eq:genRespb} in the presence of both SOC of several types and electron-electron interactions. We present a systematic comparison of \textit{both} types of cross-responses ($\sigma^{\rm ME}$ and $\chi^{\rm JB}$) in single-valley 2DEGs with Rashba or Dresselhaus SOC, such as commonly encountered in semiconductor heterostructures, as well as in multi-valley Dirac systems away from the charge-neutrality point with proximity-induced SOC of both Rashba and valley-Zeeman (VZ) types, such as in graphene on a TMD substrate \cite{Wang2015,morpurgo:prx}. The main difference between these two  systems is not the electron energy spectrum (parabolic vs. linear), but the dimensionality of the Hilbert space, which allows for mixed spin-valley excitations in multi-valley systems but not in single-valley ones. For brevity, nevertheless, we will dub the first class of systems as ``2DEG'' and the second one as ``Dirac''.

Our results can be divided in two categories. The first one includes theoretical predictions for resonance lineshapes in cross-responses, while the second one contains proposals for utilizing cross-response resonances in spintronic devices. In the first category, we show that, both in single- and multi-valley systems, CSMs corresponding to in-plane fluctuations of the spin lead to resonant versions of the Edelstein and the inverse-Edelstein/spin-galvanic effects. This in itself is not surprising. In particular, the resonance in the Edelstein conductivity, $\sigma^{\rm ME}$, in the non-interacting limit of a single-valley 2DEG, was already pointed out in Refs. ~\cite{Erlingsson2005,Bryksin2006} (although in a different context, centered around spin accumulation effects).  What is special for a multi-valley system is that the inter-valley electron-electron interaction splits the modes into two. We analyze the spectral weights of the two modes in the cross-responses as functions of the various parameters of the model and show that the low-energy mode carries most of the spectral weight. Next, we point out that cross-response arises only due to SOC that couples the momentum to spin. Thus, both Rasbha and Dresselhaus SOCs are expected to induce cross-responses, albeit with a different tensor structure, but the VZ SOC is not, because it couples spins to valleys but not
to electron momenta.

In the second category of results, we show that the charge-to-spin conversion metric commonly
used in spintronics literature $\propto\sigma^{\rm ME}/\sigma$ also experiences a resonant enhancement due to CSMs. It should be noted that while both $\sigma^{\rm ME}$ and $\sigma$ exhibit resonances due to CSMs, the resonance in $\sigma^{\rm ME}$ is by a few orders of magnitude (depending on the extent of disorder) stronger than in $\sigma$, and thus $\sigma^{\rm ME}/\sigma$ is enhanced as well. Finally, we propose a protocol for 
spin-pumping in graphene with proximity-induced SOC, based on CSMs. This protocol allows one to control the direction of pumped spins.

The rest of the paper is organized as follows. In Sec.~\ref{Sec:KinEq}, we discuss the formalism of the Landau kinetic equation for a Fermi-liquid with SOC. The electrically driven direct and cross-responses in both single and multi(two)-valley systems with Rashba SOC are discussed in Sec. \ref{Sec:Res_Edelstein}, while the magnetically driven responses are discussed in Sec. \ref{Sec:Res_InvEdel}. The same effects for a single-valley system with Dresselhaus SOC are discussed in Sec. \ref{Sec:Dresselhaus}. In Secs. \ref{subsec:2D interaction} and \ref{subsec:Dirac interaction}, we analyze the effect of electron-electron interactions (\textit{eei}).  In Sec. \ref{sec:analysis}, we discuss the spectral weights of interaction-split resonances in a multi-valley system. In Sec. \ref{Sec:Exp},  we discuss potential applications of our results in spintronics. Our conclusions are given in Sec.~\ref{Sec:Conclusion}. Appendices \ref{App2DEG} and \ref{AppGRa} provide details of solving the equations of motion for single- and two-valley systems.

\section{The Landau kinetic equation}\label{Sec:KinEq}
Following the method summarized in Ref. \cite{Maslov2022}, we employ the Landau kinetic equation for a Fermi liquid \cite{landau1980} in the following form (here and thereafter we set $\hbar=1$):
\bea\label{Eq:LKE}
\partial_t\hat n+i[\hat \epsilon,\hat n] + \frac 12\{\partial_{\br}\hat n,\partial_{\bk}\hat \epsilon\}- \frac 12\{\partial_{\bk}\hat n,\partial_{\br}\hat \epsilon\}=\hat I_{\rm coll},
\eea
where $\hat \epsilon\equiv \hat  \epsilon(\bk,t)$ is the quasiparticle energy and $\hat n\equiv \hat n(\bk,t)$ is the non-equilibrium distribution function. Both $\hat\epsilon$ and $\hat n$ are matrices of the corresponding dimensionality in the spin and valley subspaces, which are parameterized as
\bse
\bea
\hat \epsilon\equiv \hat  \epsilon(\bk,t)&=&\hat H_0(\bk)+\hat H_{\rm SOC}(\bk) 
+\hat \epsilon_{\rm LF}(\bk,t) 
+ \delta \hat\epsilon_{\rm ext}(t),\label{eq:definitions_eps}\\
\hat n\equiv \hat n(\bk,t)&=&\hat n_0(\bk) +\hat n_{\rm SOC}(\bk)
+\delta \hat n(\bk,t).\label{eq:definitions_n}
\eea
\ese
In Eq.~\eqref{eq:definitions_eps},
$\hat H_0(\bk)$ is 
the single-particle Hamiltonian in the absence of SOC, $\hat H_{\rm SOC}(\bk)$ describes the SOC,
\beq\label{eq:LF}
\hat\epsilon_{\rm LF}(\bk,t)=\int_{\bk'} {\rm Tr} [\hat F(\bk,\bk')\delta \hat n(\bk',t)] \eeq
describes the effect of interaction between quasiparticles via the Landau functional (LF) $\hat F(\bk,\bk')$ with $\int_{\bk}\equiv \int d^2k/(2\pi)^2$, and $\delta\hat \e_{\rm ext}(t)$ is a correction to the energy due to the driving oscillatory electric and magnetic fields. Explicit forms of $\hat{H}_0(\bk)$ and $\hat{H}_{\text{SOC}}(\bk)$ will be presented later in Sec.~\ref{Sec:Res_Edelstein}. The Landau functional is taken to be the same as when SOC is absent. This means that SOC is assumed to be weak, such that $\lambda_{\rm SOC}/\mu\ll1$ (where $\mu$ is the chemical potential and $\lambda_{\rm SOC}$ is the energy splitting of states at the Fermi level), and is treated perturbatively. The last condition automatically implies that we can only consider the situation when both spin-split subbands are occupied.

Next, in Eq.~\eqref{eq:definitions_n}, $\hat n_0(\bk)$ is the equilibrium electron distribution, $\hat n_{\rm SOC}(\bk)=n'_0\hat H_{\rm SOC}(\bk)$ is the correction due to SOC with $n'_0\equiv \partial_{\epsilon} n_0$, and  $\delta \hat n(\bk,t)$ is the non-equilibrium part of $\hat n$. Note that the deviation from equilibrium enter both directly, via $\delta \hat n(\bk,t)$, and indirectly, via the corresponding change in the quasiparticle energy, as specified by Eq.~\eqref{eq:LF}. 

Finally, the collision term on the right-hand side of Eq.~\eqref{Eq:LKE} accounts for relaxation processes. 
A microscopic description of relaxation processes is beyond the scope of this paper. At the phenomenological level, $\hat I_{\rm coll}$ can be decomposed into the parts corresponding to relaxation of different degrees of freedom, i.e., charge, in-plane and out-of-plane components of spins, and valley polarization. These degrees of freedom will be made explicit in \cref{Sec:Res_Edelstein}. For each component, one can introduce a phenomenological relaxation time \cite{RAKITSKII2025}. We will assume that our resonances lie in the ballistic regime, where $\lambda_{\rm SOC}\tau_{\min}\gg1$ ($\tau_{\min}$ being the shortest relaxation time). Thus, the net effect of $\hat I_{\rm coll}$ is only to introduce an effective broadening of the resonances. To avoid dealing with the multitude of phenomenological parameters, we set all the relaxation times to be equal to some effective relaxation time $\tau$ and model the collision integral as $\hat I_{\rm coll}=-\delta\hat n(\bk)/\tau$. Microscopically, such a case is encountered--up to numerical coefficients--in the ballistic limit of the D'yakonov-Perel' spin-relaxation mechanism due to scattering by short-range impurities in the presence of Rashba and/or Dresselhaus SOC \cite{DYAKONOV1972}. In general, the scattering time $\tau$ may depend on the electron energy.  However, since we are interested only in low-energy excitations near the Fermi surface, $\tau$ can be evaluated right at the Fermi energy. Unlike some other SOC-related phenomena, e.g., Hanle effect \cite{Golub2025}, CSMs are not sensitive to the particle-hole asymmetry. As a result, the energy dependence of $\tau$ is not relevant for their damping. We caution the reader that, ss a consequence of choosing to work in the ballistic regime, the limit of $\Omega=0$ cannot be taken in the results presented in this paper.

The driving $E$ and $B$-fields are assumed to be weak and spatially uniform, such that the kinetic equation can be linearized in $\delta\hat n$ and $\delta\hat \e_{\rm ext}(t)$. In accord with an assumption of weak SOC, it will also be linearized in $\hat H_{\rm SOC}$. In this regard, a term $\propto\hat H_{\rm SOC}\hat \e_{\rm ext}$ is allowed as it is still linear in its constituents. If the external driving force is due to the $E$-field, the linearized kinetic equation takes the form\footnote{Here, we ignore the term $\propto n''_0\hat H_{\rm SOC}\mathbf{v}_{\bk}\cdot \mathbf E_0$, as it does not affect the resonant behavior that we wish to model.}: 
\beq\label{eq:EdriveGenE}
\partial_t\delta\hat n + i[\hat H_{\rm SOC},\delta \hat n]-in'_0[\hat H_{\rm SOC},\int_{\bk'}\hat F_{\bk\bk'}\delta\hat n_{\bk'}]=e n'_0(\hat {\mathbf v}_{\bk}+\hat{\mathbf v}_{\rm SOC})\cdot\mathbf E_0 e^{-i\Omega t}+\hat I_{\rm coll},
\eeq
where $\mathbf{E}_0e^{-i\Omega t}$ is the applied oscillatory $E$-field which has only in-plane components, and $
\hat{\bv}_\bk \equiv \partial\hat{H}_0/\partial\bk$ and $\hat{\bv}_\text{SOC} \equiv \partial\hat{H}_\text{SOC}/\partial\bk$ are the parts of the velocity operator with and without SOC, respectively.
The coupling of the $E$-field to electron spins, i.e., the EDSR effect, occurs via the $e\hat n_0' \hat{\mathbf v}_{\rm SOC}\cdot\mathbf E_0 e^{-i\Omega t}$ term on the right-hand side of Eq.~\eqref{eq:EdriveGenE}. For the $B$-driven case, the equation takes the form\footnote{Since, ${\bf B}_0$ is assumed to be weak, we only model the coupling of the oscillatory $B$-field to the spin and ignore the orbital effect.}
\beq\label{eq:EdriveGenB}
\partial_t\delta\hat n + i[\hat H_{\rm SOC},\delta \hat n]-in'_0[\hat H_{\rm SOC},\int_{\bk'}\hat F_{\bk\bk'}\delta\hat n_{\bk'}]=-in'_0\frac{g\mu_{\rm{B}}}{2}[\mathbf B_0\cdot\hat{{\mathbf s}}+B_z\hat s_z,\hat H_{\rm SOC}]e^{-i\Omega t} +\hat I_{\rm coll},
\eeq
where $(\mathbf{B}_0,B_z)e^{-i\Omega t}$ is the applied oscillatory $B$-field. Throughout the paper, we will use bold symbols to denote in-plane vectors, while the $z$-component will be labeled explicitly. Note that a spatially-uniform but oscillatory $B$-field does not induce magnetization in the absence of SOC because the spin susceptibility vanishes in this limit as dictated by spin conservation. Accordingly, the driving term in the right-hand side of Eq.~\eqref{eq:EdriveGenB} vanishes for $\hat H_{\rm SOC}=0$.

After solving for $\delta\hat n$, we will be interested in calculating two observables: the electric current density, as deduced from the continuity equation, and magnetization. These are given by \cite{nozieres,physkin}
\bse
\bea
{\bf J}&=&-\frac{e}{2}\int_{\bk}{\rm Tr}\left[\{\partial_\bk(\hat H_0+\hat H_{\rm SOC}),(\delta\hat n-\hat n'_0\hat \e_{\rm LF})\}\right],\label{eq:responsesJ}\\
{\bf M}&=&-\frac{g\mu_{\rm{B}}}{2}\int_\bk{\rm Tr}[\hat{\mathbf s}\delta\hat n],~~M_z=-\frac{g\mu_{\rm{B}}}{2}\int_\bk{\rm Tr}[\hat{s}^z\delta\hat n],\label{eq:responsesM}
\eea 
\ese
where $-e$ is the electron charge, $g$ is the effective Land\'e $g$-factor, $\mu_{\rm{B}}$ is the Bohr magneton, and $\hat s^\alpha$ with $\alpha\in\{x,y,z\}$ are spin matrices. The traces are taken over both the spin and valley subspaces. All the parameters of the single-particle Hamiltonians in the above equations--$g$, $v_{\rm SOC}$, etc.--are assumed to include Fermi-liquid renormalizations. To simplify notations, we will use the same symbols for bare and renormalized quantities.

\section{Electric-field driving: Electric-dipole spin resonance and the resonant Edelstein effect}
\label{Sec:Res_Edelstein}
Both EDSR and the resonant Edelstein effect are driven by the oscillatory $E$-field. We first consider a single-valley system with Rashba SOC, in which case
\bea\label{Eq:2deg}
\hat H_0=\left(\dfrac{\bk^2}{2m}-\mu\right)\hat s_0
\;{\rm and}\;
\hat H_{\rm SOC}=v_{\rm R}(\bk\times\hat{\mathbf s})\cdot{\mathbf z},
\eea
where `$~\hat{~}~$' indicates a matrix in the $2\times 2$ spin space, $m$ is the effective mass, and $v_{\rm R}$ is the Rashba coupling constant with units of velocity. Correspondingly,
\bea\label{Eq:2degvel}
\hat{\mathbf v}_{\bk}=\dfrac{{\bk}}{m}\hat s_0
\;{\rm and}\;
\hat{\mathbf v}_{\rm SOC}=v_{\rm R}~(\bk_u\times\hat{\mathbf s}),
\eea
where $\bk_u$ is the unit vector along $\bk$. The kinetic equation becomes significantly easier to solve 
if the $2\times 2$ distribution function is expanded over the Pauli matrices in the ``chiral'' basis 
\cite{shekhter2005,kumar2017}:
\beq\label{eq:LKE_2deg}
\delta\hat n(\bk)=-\delta(\epsilon-\epsilon_F)u_i(\bk)\hat c^i,~~~~i\in\{0,1,2,3\},
\eeq
where ${\hat c}^0={\hat{s}}^0,~~{\hat c}^1=-{\hat{s}}^z,~~{\hat c}^2={\hat{s}}^x\cos\theta+{\hat{s}}^y\sin\theta~$, ${\hat c}^3={\hat{s}}^x\sin\theta-{\hat{s}}^y\cos\theta$, $\theta$ is azimuthal angle of $\bk$, and $-\delta(\epsilon - \epsilon_F)$ is the zero temperature limit of $\partial_\epsilon n_0$. The label $i$ denotes the degrees of freedom of the system. Due to the $\delta$-function, the functions $u_i(\bk)$ are projected onto the Fermi surface and depend only on $\theta$: $u_i(\bk)=u_i(\theta)$. Since the Rashba SOC is rotationally-symmetric, the equations for harmonics of $u_i(\theta)$ with different projections of the angular momentum onto the $\hat z$-axis decouple. It is then convenient to expand the functions $u_i(\theta)$ over a complete basis of the eigenstates of the angular momentum operator as $u_i(\theta)=\sum_{m=-\infty}^\infty u_i^{(m)}e^{im\theta}$. In terms of $u^{(m)}_i$, the components of the current density are re-written as
\bea\label{eq:J2DEG}
J_x&=&e\nu^{\rm T}_{\rm F}\left[v_{\rm{F}}\frac{u^{(-1)}_0+u^{(1)}_0}{2}+\frac{\lambda_{\rm R}}{2k_{\rm{F}}}\left(\frac{u^{(-1)}_2-u^{(1)}_2}{2i}-\frac{u^{(-1)}_3+u^{(1)}_3}{2}\right)\right],\nn\\
J_y&=&e\nu^{\rm T}_{\rm F}\left[v_{\rm{F}}\frac{u^{(-1)}_0-u^{(1)}_0}{2i}-\frac{\lambda_{\rm R}}{2k_{\rm{F}}}\left(\frac{u^{(-1)}_2+u^{(1)}_2}{2}+\frac{u^{(-1)}_3-u^{(1)}_3}{2i}\right)\right],
\eea 
where $v_{\rm{F}}$ is the Fermi velocity, $k_{\rm{F}}$ is the Fermi momentum, $\lambda_{\rm R}\equiv 2v_{\rm R} k_{\rm{F}}$ is the energy splitting due to SOC at the Fermi level, and $\nu_{\rm F}^{\rm T}=m/\pi$ is the total density of states at the Fermi level. The first term in the current density is the usual contribution from free carriers that gives the Drude response, while the second term is due to SOC. Likewise, the components of the magnetization are re-written as
\bea\label{eq:M}
M_x&=&\frac{g\mu_{\rm{B}}\nu^{\rm T}_{\rm F}}2\left(\frac{u_2^{(-1)}+u_2^{(1)}}{2}+\frac{u_3^{(-1)}-u_3^{(1)}}{2i}\right),\nn\\
M_y&=&\frac{g\mu_{\rm{B}}\nu^{\rm T}_{\rm F}}2\left(\frac{u_2^{(-1)}-u_2^{(1)}}{2i}-\frac{u_3^{(-1)}+u_3^{(1)}}{2}\right),\nn\\
M_z&=&\frac{g\mu_{\rm{B}}\nu^{\rm T}_{\rm F}}2(-u_1^{(0)}).
\eea

Next, as an example of a multi-valley system, we consider gated monolayer graphene, in which case the two valleys are centered around the $K$ and $K'$  points at the corners of the Brillouin zone. Although intrinsic SOC in graphene is very weak, substantial SOC can be induced by mounting graphene on a heavy-element substrate, e.g., TMD. It is known that SOC in this case is a combination of two types: Rashba and valley-Zeeman (VZ), the latter acting as an out-of-plane Zeeman field of polarity alternating between the $K$ and $K'$ valleys \cite{morpurgo:prx}. For definiteness, we assume that the chemical potential ($\mu$) crosses the conduction band (the upper Dirac cone). If we are interested in energies near $\mu$, one can project out the valence band (the lower Dirac cone) which leads to the following $4\times 4$ Hamiltonian:
\bea\label{eq:GrHam}
\hat H_0=(v_{\rm{F}} k-\mu) \hat \tau^0\hat s^0,~~~\hat H_{\rm SOC}=\frac{\lambda_{\rm R}}{2}\hat \tau^0(\bk_u \times \hat{\bf s})\cdot {\bf z}+\frac{\lambda_{\rm Z}}2\hat \tau^3\hat s^3.
\eea
Here, in addition to the $2\times 2$ spin matrices $\hat s$, there are also $2\times 2$ $\hat{\tau}$ matrices that act in the valley space, and $\lambda_{\rm R}$ and $\lambda_{\rm Z}$ are the Rashba and VZ spin-orbit couplings, respectively, both with units of energy. Since a gated system has a finite $k_{\rm{F}}$, 
it is also convenient to introduce a velocity $v_{\rm R}\equiv \lambda_{\rm R}/2k_{\rm{F}}$. When comparing Hamiltonian \eqref{eq:GrHam} with Eq.~\eqref{Eq:2deg}, it is worth noting that the SOC term in the former depends only on the direction of $\bk$ but not on its magnitude. This is the result of projecting out fully occupied bands in the Dirac system. Consequently, the structure of spin-orbit part of the velocity operator
\bea\label{Eq:Gvel}
\hat{\bf v}_{\bk}=v_{\rm{F}}\bk_u\hat\tau^0\hat s_0,~~~\hat{\bf v}_{\rm SOC}=v_{\rm R} (\bk_u\cdot\hat{\bf s}) (\bk_u\times\hat z)\hat\tau^0,
\eea
differs from that in Eq.~\eqref{Eq:2degvel}. Next, the non-equilibrium part of the distribution function is modeled as \cite{Raines2021,kumar2021}
\bea\label{eq:DFGr}
\delta\hat n=-\delta(\epsilon-\epsilon_F)\left[u_0\hat\tau^0\hat c^0+u_i\hat\tau^0\hat c^i+W_i\hat\tau^i\hat c^0+M_{ij}\hat\tau^i\hat c^j\right],~~~~i,j\in\{1,2,3\},
\eea
where the $u_i$ terms describe the spin-chiral sector in the same way as for 2DEG [cf.~Eq.~\eqref{eq:LKE_2deg}], the $W_i$ terms describe the valley sector, and the $M_{ij}$ terms describe coupled fluctuations in the spin-valley sector. Analogously to $u_i(\theta)$, we expand 
the valley and spin-valley terms as
\bea
W_i(\theta)=\sum_m W_i^{(m)} e^{im\theta},~M_{ij}(\theta)=\sum_m M_{ij}^{(m)} e^{im\theta}.
\eea
In the case of equally populated valleys, the valley degree of freedom is decoupled both from the other degrees of freedom, as well as from the external field. Therefore, the $W_i$ terms need not be considered.
Applying the general expression for the
electric current, Eqs.~(\ref{eq:responsesJ}),
to the two-valley case, we obtain
\bea\label{eq:JGr}
J_x&=&e\nu^{\rm T}_{\rm F}\left[v_{\rm{F}}\frac{u_0^{(-1)}+u_0^{(1)}}{2}+\frac{\lambda_{\rm R}}{2k_{\rm{F}}}\left\{\frac{u^{(-1)}_2-u^{(1)}_2}{2i}
\right\}\right],\nn\\
J_y&=&e\nu^{\rm T}_{\rm F}\left[v_{\rm{F}}\frac{u_0^{(-1)}-u_0^{(1)}}{2i}-\frac{\lambda_{\rm R}}{2k_{\rm{F}}}\left\{\frac{u^{(-1)}_2+u^{(1)}_2}{2}\right\}\right],
\eea
where 
$\nu_{\rm F}^{\rm T} \equiv 2k_{\rm{F}}/\pi v_{\rm{F}}$. The expressions for the components of the magnetization are same as in the single-valley case [Eq. (\ref{eq:M})], with $\nu_{\rm F}^{\rm T}$ corresponding now to the two-valley case. Comparing Eq.~\eqref{eq:JGr} with Eq.~(\ref{eq:J2DEG}), we see that while the Drude parts of the current ($\propto v_{\rm{F}}$) are the same for the single- and two-valley systems, the SOC induced terms are different due to a difference in the SOC parts of the velocity operators [cf. Eqs.~(\ref{Eq:2degvel}) and (\ref{Eq:Gvel})]. As we shall see below, this makes the $\Omega$-dependence of $\bf J$ in the Dirac system different from that in the 2DEG.

For the case of a single-valley 2DEG, the equations of motion for $u_i^{(m)}$ are solved in \cref{App2DEG}.
Substituting these solutions into Eq.~\eqref{eq:J2DEG}, we obtain for the $E$-field induced current
\beq\label{eq:j2D-e}
\begin{split}
\text{2DEG: }~~\bJ=  
\underbrace{
\frac{i\sigma_0}{\tau}\left[\frac1{\bar\Omega }+\frac1{\bar\Omega}\left(\frac{\lambda_{\rm R}}{4\mu}\right)^2\{2+R^{\rm 2D}_{\rm JE}(\Omega)\}\right]
}_{\equiv\sigma(\Omega)}
{\bE}_0, \,\,\,\, {\rm where}\;R_\text{JE}^{\rm 2D}(\Omega) = \frac{\lambda_{\rm R}^2}{\bar\Omega^2 - \lambda_{\rm R}^2},
\end{split}
\eeq
where the superscript ``2D'' stands for ``2DEG'', $\bar\Omega\equiv \Omega+i/\tau$, and 
\bea
\sigma_0=e^2\nu_{\rm F}^{\rm T}v_{\rm{F}}^2\tau/2\label{sigma0}
\eea 
is the \emph{dc} Drude conductivity. The prefactor of ${\bf E}_0$ is identified as the total charge conductivity $\sigma(\Omega)$. 
The first, $1/\Omega,$ term in the square brackets gives the frequency dependence of the Drude contribution, while the second term represents renormalization of the Drude weight by Rashba SOC (the additive `$2$' to $R^{\rm 2D}_{\rm JE}$), as well as the EDSR peak at $\Omega=\lambda_{\rm R}$.

Turning to a current response of two-valley Dirac system, the equations of motion for the spin and spin-valley degrees of freedom, $u^{(m)}_i$ and $M^{(m)}_{ij}$, are solved in \cref{AppGRa}. Substituting these solutions into Eq.~\eqref{eq:JGr} yields
\bea\label{eq:jGre1}
\text{Dirac:}~~\bJ&=&\underbrace{\frac{i\sigma_0}{\tau}\left[\frac1{\bar\Omega }+\frac1{\bar\Omega}\left(\frac{\bar\Omega}{2\mu}\right)^2R^{\rm Di}_{\rm JE}(\Omega)\right]}_{\sigma(\Omega)}{\bf E}_0, \,\,\,\,\text{where } R_\text{JE}^{\rm Di}(\Omega) = \frac{\lambda_{\rm R}^2}{\bar\Omega^2\underbrace{-\lambda_{\rm R}^2-\lambda_{\rm Z}^2}_{-\lambda^2_{\rm SOC}}},
\eea
where the superscript ``Di'' stands for ``Dirac", and $\sigma_0$ is the same as in Eq.~\eqref{sigma0}, with $\nu_{\rm F}^{\rm T}$  re-defined appropriately. The current responses in the 2DEG and Dirac systems have the same Drude part, which is to be expected  since this part arises from the free carriers at the Fermi surface. The functional form of the coupling to the resonance is different in the two systems, but become similar near the resonance. This difference arises from the difference in the structure of Eqs. \eqref{eq:J2DEG} and \eqref{eq:JGr}. We also see from Eq. \eqref{eq:jGre1} that, for finite $\lambda_{\rm Z}$ but $\lambda_{\rm R}=0$, the resonance part of the current response vanish for a Dirac system. This demonstrates the necessity of a Rashba type of SOC for the EDSR effect. If both Rashba and VZ types of SOC are present, the resonance occurs at frequency $\lambda_{\rm SOC}\equiv\sqrt{\lambda^2_{\rm R}+\lambda^2_{\rm Z}}$.

Now we turn to the magnetization response, starting with a 2DEG. As seen from Eqs.~(\ref{eq:J2DEG}) and (\ref{eq:M}), $\bf J$ and $\bf M$ contain the same harmonics of ${\bf u}(\theta)$. This implies that driving the system with an $E$-field produces not only a direct response, i.e., the charge current, but also a cross-response, i.e., the magnetization. A straightforward calculation  using Eq. \eqref{eq:responsesM} leads to (see \cref{App2DEG})
\bea\label{eq:m2DEGe}
\text{2DEG: }~~\bf M&=&\underbrace{-\dfrac{i\sigma^{\rm ME}_0}{\bar\Omega\tau}R^{\rm 2D}_{\rm ME}(\Omega)}_{\sigma^{\rm ME}(\Omega)}{\bf E}_0\times \hat{z},~~\text{where }R^{\rm 2D}_{\rm ME}(\Omega)=R^{\rm 2D}_{\rm JE}(\Omega),\nn\\
M_z&=&0,\eea
and
\bea
\sigma^{\rm ME}_0 \equiv \frac{g\mu_{\rm{B}}}2\frac{e\lambda_{\rm R}\nu_{\rm F}^{\rm T}\tau}{4k_{\rm{F}}}\label{sigma0ME}
\eea
is the \emph{dc} Edelstein conductivity \cite{EDELSTEIN1990}. In this context the prefactor of ${\bf E}_0$ is identified with the dynamic Edelstein conductivity $\sigma^{\rm ME}(\Omega)$, which exhibits a resonance at the same frequency $\Omega=\lambda_{\rm R}$ as the charge conductivity. Note that the induced magnetization lies in the 2DEG plane, transverse to the applied $E$-field, and that the prefactor of its resonant part ($\sigma_0^{\rm ME}/\tau$) is independent of the relaxation time $\tau$ since $\sigma^{\rm ME}_0\propto\tau$.

For the Dirac case, we need $u^{(m)}_i$ and $M^{(m)}_{ij}$ components which are found by solving the kinetic equation (\cref{AppGRa}). To calculate the magnetization, we again use Eq.~\eqref{eq:responsesM} which also leads to Eq. \eqref{eq:M}. This gives 
\bse
\bea\label{eq:jGre12}
\text{Dirac:}~~{\bf M}&=&\underbrace{-\dfrac{i\sigma^{\rm ME}_0}{\bar\Omega\tau}\left(\frac{\bar\Omega}{\lambda_{\rm R}}\right)^2R^{\rm Di}_{\rm ME}(\Omega)}_{\sigma^{\rm ME}(\Omega)} {\bf E}_0\times\hat z,\;
R^{\rm Di}_{\rm ME}(\Omega)=R^{\rm Di}_{\rm JE}(\Omega),\\
M_z&=&0,
\eea
\ese
where $\sigma^{\rm ME}_0$ is the same as in Eq. \eqref{sigma0ME}, with $\nu_{\rm F}^{\rm T}$ re-defined appropriately. Observe that {\bf M} in Eq.~\eqref{eq:jGre12} vanishes if $\lambda_{\rm R}=0$, irrespective of $\lambda_{\rm Z}$.

\section{Magnetic field driving: Chiral-spin resonances and the resonant inverse Edelstein effect}\label{Sec:Res_InvEdel}
Both chiral-spin resonances and the resonant inverse Edelstein effect are driven by the oscillatory $B$-field. Solving the kinetic equation in the presence of magnetic driving, we find (see \cref{App2DEG,AppGRa})
\bea\label{eq:mag2DEGb}
\text{2DEG:}~\bf M&=&\underbrace{-\frac{\chi_0}{2}R^{\rm 2D}_{\rm MB}(\Omega)}_{\chi_{\parallel}(\Omega)}{\bf B}_0,\,\,\,\, R_\text{MB}^{\rm 2D}(\Omega) = R_\text{JB}^{\rm 2D}(\Omega), \nn\\
M_z&=&\underbrace{-\chi_0 R^{\rm 2D}_{\rm MB}(\Omega)}_{\chi_{\perp}(\Omega)}B_z;\nn\\
\text{Dirac: }~\bf M&=&\underbrace{-\frac{\chi_0}2R^{\parallel \rm Di}_{\rm MB}(\Omega)}_{\chi_{\parallel}(\Omega)} {\bf B}_0, \,\,\,\,\,\, 
R^{\parallel \rm Di}_{\rm MB}(\Omega)=\frac{\lambda_{\rm R}^2+2\lambda_{\rm Z}^2}{\bar\Omega^2-\lambda_{\rm SOC}^2}, \nn\\
M_z&=&\underbrace{-\chi_0R^{\perp \rm Di}_{\rm MB}(\Omega)}_{\chi_{\perp}(\Omega)}B_z, \,\,\,\,\,\, R^{\perp \rm Di}_{\rm MB}(\Omega)= \frac{\lambda_{\rm R}^2}{\bar\Omega^2-\lambda_{\rm SOC}^2},
\eea
where
\bea
\chi_0\equiv \frac14 g^2\mu_{\rm{B}}^2\nu^{\rm T}_{\rm F}\label{chi0}
\eea
is the static spin susceptibility in the absence of SOC and 
\bea
\chi_0^{\rm JB}\equiv \frac{\sigma^{\rm ME}_0}{\tau}
=\frac{g\mu_{\rm{B}}}{2}\frac{e\lambda_{\rm R}\nu_{\rm F}^{\rm T}}{4k_{\rm{F}}}\label{chi0JB}.
\eea
Once again, $\nu_{\rm F}^T$ is appropriately re-defined. While the static spin susceptibility remains isotropic even in the presence of SOC (provided that both Rashba subbands are occupied) \cite{zak:2010}, the \emph{dynamic} susceptibility is anisotropic, such that its in-plane  component ($\chi_\parallel$) is twice smaller than the out-of-plane one ($\chi_\perp$).

We already know from Eqs. (\ref{eq:J2DEG}), (\ref{eq:M}) and \eqref{eq:JGr} that 
whatever induces ${\bf M}$ must also induce ${\bf J}$. For the case of $B$-field driving, the induced charge current represents the resonant inverse Edelstein effect. Note that the current is induced only by the in-plane component of the $B$-field and in direction perpendicular to ${\bf B}_0$, and that, similar to the case of Edelstein effect, the prefactor of its resonant part is independent of $\tau$. For $B$-field driving, we obtain
\bse
\bea
\text{2DEG: }~\bf{J}&=&\underbrace{-\chi_0^{\rm JB} R^{\rm 2D}_{\rm JB}(\Omega)}_{\chi^{\rm JB}(\Omega)}{\bf B}_0\times\hat{z}, \,\,\,\, R_\text{JB}^{\rm 2D}(\Omega) = \frac{\lambda_{\rm R}^2}{\bar\Omega^2-\lambda_{\rm R}^2},\label{eq:BDrive}\\
\text{Dirac: }~\bf J&=&-\chi_0^{\rm JB}R^{\rm Di}_{\rm JB}~{\bf B}_0\times\hat z, \,\,\,\,\,\, R^{\rm Di}_{\rm JB}(\Omega)=\frac{\lambda_{\rm R}^2+\lambda_{\rm Z}^2}{\bar\Omega^2-\lambda_{\rm SOC}^2},\label{eq:BDrive_b}
\eea
\ese
where $\chi_0$ and $\chi_0^{\rm JB}$ are the same as in Eqs.~\eqref{chi0} and \eqref{chi0JB}, respectively, with $\nu_{\rm F}^{\rm T}$ appropriately defined. Thus, the cross-responses persist in multi-valley systems with the property that it vanishes when $\lambda_{\rm R}=0$ (irrespective of $\lambda_{\rm Z}$). We also remind the reader that while these results are valid for $\Omega$ around the resonance, our model form of the collision integral does not allow us to extend  Eqs.~\eqref{eq:BDrive} and \eqref{eq:BDrive_b} down to the static limit, where the induced current must vanish\cite{shen2014}.

\begin{figure*}[htp]
\centering
\includegraphics[width=0.9\linewidth]{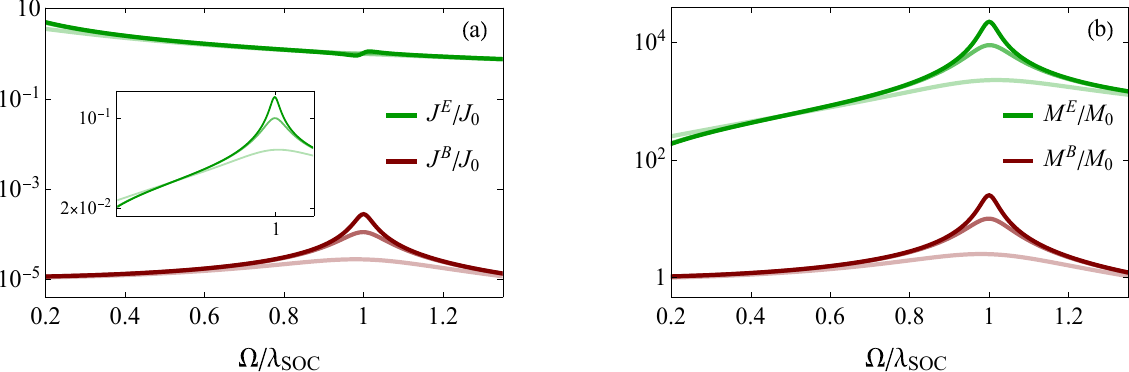}
\caption{\textbf{Cross responses vs direct responses}. (a) Direct ($J^E$) and cross-response ($J^B$) parts of the induced current normalized to $J_0=e^2v_{\rm F}^2\nu_{\rm F}^{\rm T}E_0/2\lambda_{\rm SOC}$ in a two-valley Dirac system. The inset shows the $J^E$ response without the Drude tail. (b) Direct ($M^B$) and cross-response ($M^E)$ parts of the induced magnetization, normalized to $M_0=\chi_0B_0/2$, which is the magnetization of free electrons in a static $B$-field of amplitude $B_0$. The responses are plotted for $\tau\lambda_{\rm SOC}=50,20,5$ (from dark curves to light ones), $\lambda_{\rm Z}=0.5\lambda_{\rm R}$, and $\mu=5\lambda_{\rm R}$.}\label{fig:1}
\end{figure*}

We now examine the case of a normally incident electromagnetic wave, when the amplitudes of the electric and magnetic fields are related by $E_0=cB_0$. In Figs.~\ref{fig:1}(a) and (b),  we show the magnitudes of the $E$-driven and $B$-driven current and magnetization for a two-valley system, denoted by $J^E=\sigma(\Omega)E_0$, $J^{B}=\chi^{\rm JB}(\Omega) E_0/c$, $M^E=\sigma^{\rm ME} cB_0$, and $M^B=\chi_{||}(\Omega)B_0$, respectively. In the non-interacting limit, the results for a single-valley system are similar. Several features need to be pointed out. i) Electrically driven responses (both direct and cross) are orders of magnitude stronger than magnetically driven ones,  even away from the resonance. ii) Although the total current induced by the $E$-field is much larger than that induced by the $B$-field, it is largely dominated by a featureless Drude background, making the EDSR peak barely discernible--even for a large quality factor, e.g., $\lambda_{\rm SO}\tau=50$. On the other hand, the resonance in a much smaller $B$-induced current is well resolved. iii) For magnetization, the resonances are well resolved in both direct ($B$-induced) and cross ($E$-induced) cases, but the signal is stronger in the cross-response. This remains true as long as $\sigma^{\rm ME}(\Omega)c\gg\chi^{\rm MB}(\Omega)$, which translates into $\lambda_{\rm R}/\Omega\gg gk_F/k_C\sim10^{-3}$, where $k_C=m_ec$ is the Compton wavevector, with $m_e$ and $c$ being the bare electron mass and speed of light in vacuum, respectively. This condition is easily satisfied near and below resonance, but not at higher frequencies. Observe that magnetization cross-response has an additional enhancement near $\Omega\approx\lambda_{\rm SOC}$ of at least an order of magnitude. This is due to the CSMs of the system.

\paragraph*{Remark:} Before proceeding further, we need to make the following observation. If only Rashba SOC is present in the Dirac system, then the magneto-electric response for both 2DEG and Dirac systems near resonance and for normal incidence can be written in a unified form, namely,
\begin{subequations}
\bea
\mathbf M &=& -\frac{\chi_0}{2}\left(R(\Omega)
~
\mathbf B_0+i\frac2g\frac{k_C}{k_F}R(\Omega){\rm sgn}(\lambda_R)~\frac{\mathbf E_0}c\times\hat z\right),
\label{eq:genResp2}\\
\mathbf J\!\!&=&\!\!\frac{i\sigma_0}
{\Omega\tau}\left(\!\mathbf E_0+s^2R(\Omega)\mathbf E_0+i\frac{g}{2}\frac{k_F}{k_C}s^2R(\Omega){\rm sgn}(\lambda_R)~c\mathbf B_0\times \hat z\!\right)\label{eq:genResp3}
\eea
\end{subequations}
where $s\equiv v_{\rm R}/v_{\rm F}$ and $R(\Omega)=\lambda_{\rm R}^2/(\Omega^2-\lambda_{\rm R}^2)$. The first term in Eq.~\eqref{eq:genResp2} represents CSR, while the second one represents the resonant Edelstein effect, which is phase-shifted by $\pi/2$ with respect to the $E$-field. For an electromagnetic wave with $E_0=cB_0$, the second term is larger than the first one by a factor of $k_C/k_F$, which is $\sim 10^3-10^4$ for a typical semiconductor heterostructure. The first term in Eq.~\eqref{eq:genResp3} is the Drude part of the current, while the second term represents EDSR. The amplitude of the term is proportional to $s^2$, which is typically small ($\lesssim10^{-3}-10^{-4}$), and thus EDSR can be discerned against the Drude part only if the quality factor  is very large~\cite{maiti2015a}. Finally, the third term in Eq.~\eqref{eq:genResp3} represents the resonant inverse Edelstein effect, whose amplitude is suppressed by $k_F/k_C$ when compared to the EDSR.

\section{Cross responses in the presence of Dresselhaus spin-orbit coupling}\label{Sec:Dresselhaus}
The existence  of cross-responses is not unique to Rashba SOC. Any perturbation that mixes spin and momentum will lead to similar effects but with a different tensor structure of the response functions. As another example, we consider Dresselhaus type of SOC, encountered on surfaces of crystals without a bulk inversion  center \cite{Dresselhaus1955,Dyakonov:book}. For a 2DEG on the (001) surface plane, the Hamiltonian of Dresselhaus SOC reads
\bea\label{Eq:2deg-Dresselhaus}
\hat H_{\rm SOC}=v_{\rm D}(\hat s_x k_x-\hat s_yk_y),
\eea
with $v_{\rm D}$ being the Dresselhaus coupling constant. Correspondingly, the spin-orbit part of the velocity becomes
\bea\label{Eq:2degvel-D}
\hat{\bf v}_{\rm SOC}=v_{\rm D}(k_{ux}\hat s_x\hat{\bf x}-k_{uy}\hat s_y\hat{\bf y}).
\eea
One can now proceed along the same lines as for the Rashba case. For electric-field driving we find:
\bea\label{eq:jm2dE-D}
\bf J&=&\frac{i\sigma_0}{\tau}\left[\frac1{\bar\Omega }+\frac1{\bar\Omega}\left(\frac{\lambda_{\rm D}}{4\mu}\right)^2D_{\rm JE}^{\rm 2D}(\Omega)\right]{\bf E}_0,\nn\\
M_x&=&\dfrac{i\sigma^{\rm ME}_0}{\bar\Omega\tau}D_{\rm ME}^{\rm 2D}(\Omega)E_x,\;
M_y=-\dfrac{i\sigma^{\rm ME}_0}{\bar\Omega\tau}D_{\rm ME}^{\rm 2D}(\Omega)E_y.\eea
where $D_{\rm JE}^{\rm 2D}(\Omega)=D_{\rm ME}^{\rm 2D}(\Omega)$ is same as $R_{\rm JE}^{\rm 2D}(\Omega)=R_{\rm ME}^{\rm 2D}(\Omega)$ in Eq. \eqref{eq:j2D-e}, but with $v_{\rm R}\rightarrow v_{\rm D}$, which results in $\lambda_{\rm R}\rightarrow\lambda_{\rm D}$. The same applies to the definition of $\sigma^{\rm ME}_0$ in Eq.~\eqref{sigma0ME}. For $B$-field driving, we get
\bea\label{eq:jm2DB-D}
J_x&=&-\frac{\sigma_0^{\rm ME}}{\tau}D_{\rm JB}^{\rm 2D}(\Omega)B_x,\;
J_y=\underbrace{\frac{\sigma_0^{\rm ME}}{\tau}D_{\rm JB}^{\rm 2D}(\Omega)}_{\chi_{\rm JB}}B_y,\nn\\
\bf M&=&\underbrace{-\frac{\chi_0}2D_{\rm MB}^{\rm 2D}(\Omega)}_{\chi_\parallel}{\bf B}_0,\;
M_z=\underbrace{-\chi_0D_{\rm MB}^{\rm 2D}(\Omega)}_{\chi_\perp}B_z,
\eea
where $\chi_0$ is same as in Eq.~\eqref{chi0} and
$D_{\rm JB}^{\rm 2D}(\Omega)
=D_{\rm MB}^{\rm 2D}(\Omega)$ are also same as $R_{\rm JB}^{\rm 2D}(\Omega)=R_{\rm MB}^{\rm 2D}(\Omega)$ [Eq.~\eqref{eq:mag2DEGb}], but, again, with $v_{\rm R}\rightarrow v_{\rm D}$.

In a compact form, the cross-response tensors for pure Rashba and pure Dresselhaus cases are:
\bea\label{eq:TensorStructures}
\sigma^{\rm ME}_{ab}(\Omega)&=&\begin{cases}
    \frac{-i\sigma_0^{\rm ME}}{\bar\Omega\tau}R^{\rm 2D}_{\rm ME}(\Omega)\begin{pmatrix}
        0&1\\-1&0
    \end{pmatrix},~\text{for Rashba SOC},\\
\frac{-i\sigma_0^{\rm ME}}{\bar\Omega\tau}D^{\rm 2D}_{\rm ME}(\Omega)\begin{pmatrix}
        -1&0\\0&1
    \end{pmatrix},~\text{for Dresselhaus SOC}.
\end{cases}\nn\\
\chi^{\rm JB}_{ab}(\Omega)&=&\begin{cases}
    -\chi_0^{\rm JB}R^{\rm 2D}_{\rm JB}(\Omega)\begin{pmatrix}
        0&1\\-1&0
    \end{pmatrix},~\text{for Rashba SOC},\\
-\chi_0^{\rm JB}D^{\rm 2D}_{\rm JB}(\Omega)\begin{pmatrix}
        1&0\\0&-1
    \end{pmatrix},~\text{for Dresselhaus SOC}.
\end{cases}
\eea
A characteristic difference in the tensor structure of the cross-response functions in the two cases allows for an unambiguous identification of  the dominant type of SOC in a given material. This is to be contrasted with the tensor structure of direct response functions, i.e., conductivity and spin susceptibility, which are diagonal for both Rashba and Dresselhaus cases. The difference in the tensor structure of cross-response functions could be exploited in transport measurements to extract different components of SOC. The anisotropic aspect of the tensor has been explored in the context of both the Edelstein effect and its inverse in the static and diffusive regime~\cite{Ganichev2004,Giglberger2007,Raichev2007,Shen2014b,Gorini2017,Sheikhabadi2018,Tao2021,Suzuki2023,Dey2024}; in the present work, we have explored its dynamical version in the ballistic regime.

\section{Effect of electron-electron interaction on resonant cross-responses\label{Sec:Interactions}}
Electron-electron interaction (\emph{eei}) affects the dynamic response of a system with SOC in a number of ways. The most important one is that it endows electron spins with rigidity with respect to finite-$q$ perturbations. As a result, the various resonances both in the direct and cross sectors become dispersive modes of a FL with SOC \cite{ashrafi2012,Perez2016,maiti2017}. At $q=0$ (which is the subject of this paper), \emph{eei} renormalizes the resonance frequencies. For weak SOC, such renormalizations can be expressed in terms of the Landau parameters of the underlying Fermi liquid in the absence of SOC \cite{shekhter2005,kumar2017,kumar2021}. In addition, \emph{eei} is responsible for splitting of both ESR and EDSR peaks in a two-valley Dirac system \cite{kumar2021}. Finally, \emph{eei} leads to damping of the resonances even in the absence of disorder \cite{maiti2015b}. 

While the effects of \emph{eei} on direct responses (ESR and EDSR) have been studied extensively in the past (see review \cite{Maslov2022} and references therein), here we extend the analysis to cross-responses. To do so, we come back to Eqs.~(\ref{Eq:LKE}--\ref{eq:LF}) and restore the Landau interaction functional, $\hat\e_{\rm LF}$. Since the Landau functionals are different for a single- and multi-valley cases, we will treat the two separately, in Secs.~\ref{subsec:2D interaction} and \ref{subsec:Dirac interaction}, respectively. Furthermore, \emph{eei} also renormalizes the parameters of the free Hamiltonian, i.e., $v_{F} \to v_{F}^*$, $v_{\rm SOC} \to v_{\rm SOC}^*$ and $g \to g^*$. However, as stated earlier, we shall suppress the asterisks for brevity with the understanding that all parameters in what follows are already renormalized.

\subsection{Single-valley system}\label{subsec:2D interaction}
We start with the standard form of the Landau function for a single-valley system in the absence of both SOC and $B$-field \cite{landau1980}
\bea\label{eq:ALF}
	\nu_{\rm F}^{\rm T} F_{\varsigma_1,\varsigma_2;\varsigma_3,\varsigma_4}(\theta,\theta')=F^S(\theta,\theta')\delta_{\varsigma_1\varsigma_3}\delta_{\varsigma_2\varsigma_4}+F^A(\theta,\theta')\vec{s}_{\varsigma_1\varsigma_3}\cdot\vec{s}_{\varsigma_2\varsigma_4},
\eea
where $\varsigma_i$ label the spin components, $F^{S/A}$ are the symmetric and antisymmetric parts describing the interaction in the charge and spin sectors, respectively, and $\vec{a}$ denotes a vector $a$ with three Cartesian components. The angles $\theta,\theta'$ are the azimuthal angles of momenta $\bk,\bk'$ in Eq.~\eqref{eq:LF} with magnitude $k_{\rm F}$;
in a rotationally-invariant system, the Landau function depends only on $\theta-\theta'$. The components of the Landau function can be decomposed into angular harmonics according to $X(\theta,\theta')=\sum_m X_me^{i m(\theta-\theta')}$, where $X\in\{F^S,F^A\}$ and $m$ is the $z$-component  of the angular momentum. It will also be convenient to introduce the following combinations of the Landau parameters:
\bea\label{eq:fs}
&f^{S,(m)}=1+F^S_m,~~~f^{S,(m)}_{+}=1+\frac{F^S_{m-1}+F^S_{m+1}}{2},~~~f^{S,(m)}_{-}=\frac{F^S_{m-1}-F^S_{m+1}}{2};&\nonumber\\
&f^{(m)}=1+F^A_m,~~~f^{(m)}_{+}=1+\frac{F^A_{m-1}+F^A_{m+1}}{2},~~~f^{(m)}_{-}=\frac{F^A_{m-1}-F^A_{m+1}}{2}.&
\eea
In the presence of \emph{eei}, one needs to use the full form of the electric current with the Landau functional taken into account, as given by Eq.~(\ref{eq:responsesJ}). The (temporal Fourier transform of) current density is then evaluated as
\bea\label{eq:j-int-2d}
J_x&=&e\nu^{\rm T}_{\rm F}\left[v_{\rm{F}}f^{S,(1)}\frac{u^{(-1)}_0+u^{(1)}_0}{2}+\frac{\lambda_{\rm R}}{2k_{\rm{F}}}f^{(0)} \left( \frac{u^{(-1)}_2-u^{(1)}_2}{2i}-\frac{u^{(-1)}_3+u^{(1)}_3}{2} \right) \right],\nn\\
J_y&=&e\nu^{\rm T}_{\rm F}\left[v_{\rm{F}}f^{S,(1)}\frac{u^{(-1)}_0-u^{(1)}_0}{2i}-\frac{\lambda_{\rm R}}{2k_{\rm{F}}}f^{(0)} \left( \frac{u^{(-1)}_2+u^{(1)}_2}{2}+\frac{u^{(-1)}_3-u^{(1)}_3}{2i} \right) \right].
\eea
The definition of magnetization in Eq.~(\ref{eq:responsesM}) does not involve the Landau function, hence its explicit form remains the same as in the non-interacting case, Eq.~\eqref{eq:M} (modulo ``silent" renormalization of the Hamiltonian parameters).

Solving for the components of $u$ (\cref{App2DEG}), we obtain the following responses for electric-field driving
\bea\label{eq:2DEG-int}
\bf J&=&\underbrace{\frac{i\sigma_0}{\tau}\left[\frac{f^{S,(1)}}{\bar\Omega }+\frac{f^{(0)}}{\bar\Omega}\left(\frac{\lambda_{\rm R}}{4\mu}\right)^2\{2+R^{\rm 2D}_{\rm JE}(\Omega)\}\right]}_{\sigma(\Omega)}{\bf E}_0,\;R^{\rm 2D}_{\rm JE}(\Omega)=\dfrac{(f_+^{(1)}+f_-^{(1)})f^{(1)}\lambda_{\rm R}^2 }{\bar\Omega^2-f_+^{(1)}f^{(1)}\lambda_{\rm R}^2},\nn\\
\bf M&=&\underbrace{\dfrac{-i\sigma^{\rm ME}_0}{\bar\Omega\tau}R^{\rm 2D}_{\rm ME}(\Omega)}_{\sigma^{\rm ME}(\Omega)}({\bf E}_0\times\hat{\bf z} ),\;M_z=0,\;R^{\rm 2D}_{\rm ME}(\Omega)=R^{\rm 2D}_{\rm JE}(\Omega),
\eea
and for $B$-field driving 
\bea\label{eq:2DEGbc}
\bf{J}&=&\underbrace{-\chi_0^{\rm JB}R^{\rm 2D}_{\rm JB}(\Omega)}_{\chi^{\rm JB}(\Omega)}{\bf B}_0\times\hat{z},~~R^{\rm 2D}_{\rm JB}(\Omega)=\dfrac{f^{(0)}f^{(1)}\lambda_{\rm R}^2}{\bar\Omega^2-f_+^{(1)}f^{(1)}\lambda^2_R},\nn\\
\bf M&=&\underbrace{-\frac{\chi_0}{2}R^{\parallel 2D}_{\rm MB}(\Omega)}_{\chi^\parallel(\Omega)}\mathbf B,~~R^{\parallel 2D}_{\rm MB}(\Omega)=\dfrac{f^{(1)}\lambda_{\rm R}^2}{\bar\Omega^2-f_+^{(1)}f^{(1)}\lambda^2_R},\nn\\
M_z&=&\underbrace{-\chi_0R^{\perp 2D}_{\rm MB}(\Omega)}_{\chi^\perp(\Omega)}B_z,~~R^{\perp 2D}_{\rm MB}(\Omega)=\dfrac{f_+^{(0)}\lambda_{\rm R}^2}{\bar\Omega^2-f^{(0)}f_+^{(0)}\lambda^2_R}.
\eea
As stated, the resonance frequencies are renormalized by \emph{eei}, but differently so for the in- and out-of-plane components of $M$ (for the case of $B$-field driving). While both these resonant frequencies show up in the direct response (as was also shown in Refs.  \cite{shekhter2005,kumar2017}), only the frequencies corresponding to the resonance in the in-plane components of $M$ show up in the cross-response.

\subsection{Two-valley system}\label{subsec:Dirac interaction}
The Landau function for a two-valley system can be written as \cite{Raines2021,kumar2021} 
\bea\label{eq:LFGr}
\nu_{\rm F}^{\rm T} F_{(\upsilon_1\upsilon_2;\upsilon_3\upsilon_4);(\varsigma_1\varsigma_2;\varsigma_3\varsigma_4)}(\theta,\theta')&=&F^S(\theta,\theta')\delta_{\upsilon_1\upsilon_3}\delta_{\upsilon_2\upsilon_4}\delta_{\varsigma_1\varsigma_3}\delta_{\varsigma_2\varsigma_4}\nn\\
&&+F^A(\theta,\theta')\delta_{\upsilon_1\upsilon_3}\delta_{\upsilon_2\upsilon_4}(\vec{s}_{\varsigma_1\varsigma_3}\cdot\vec{s}_{\varsigma_2\varsigma_4})
+G^A(\theta,\theta')(\vec{\tau}_{\upsilon_1\upsilon_3}\cdot\vec{\tau}_{\upsilon_2\upsilon_4})(\delta_{\varsigma_1\varsigma_3}\delta_{\varsigma_2\varsigma_4})\nn\\
&&+H(\theta,\theta')(\vec{\tau}_{\upsilon_1\upsilon_3}\cdot\vec{\tau}_{\upsilon_2\upsilon_4})(\vec{s}_{\varsigma_1\varsigma_3}\cdot\vec{s}_{\varsigma_2\varsigma_4}),
\eea
where $\upsilon_i$ label the valleys. In addition to the usual charge ($F^S$) and spin ($F^A$) components, the Landau function for the multi-valley system contains also valley ($G^A$) and spin-valley components ($H$), which account for inter-valley interactions. In Eq.~\eqref{eq:LFGr}, we neglected the difference between the in- and out-of-plane components of $G^A$ and $H$, which results from processes that swap electrons between different valleys. The matrix element of such processes is small as long as the radius of \emph{eei} is much larger than the lattice spacing, which we assume to be the case here. 

Like in the 2DEG, the components of the Landau function can be decomposed into angular harmonics according to $X(\theta,\theta')=\sum_m X_me^{i m(\theta-\theta')}$, where $X\in\{F^S,F^A,G^A,H\}$ and $m$.
After that, the expression for current density becomes
\bea\label{eq:jG-int}
J_x&=&e\nu^{\rm T}_{\rm F}\left[v_{\rm{F}}f^{S,(1)}\frac{u_0^{(-1)}+u_0^{(1)}}{2}+\frac{\lambda_{\rm R}}{2k_{\rm{F}}}\left( f^{(1)}_+\frac{u^{(-1)}_2-u^{(1)}_2}{2i}-f^{(1)}_-\frac{u^{(-1)}_3+u^{(1)}_3}{2}
\right)\right],\nn\\
J_y&=&e\nu^{\rm T}_{\rm F}\left[v_{\rm{F}}f^{S,(1)}\frac{u_0^{(-1)}-u_0^{(1)}}{2i}-\frac{\lambda_{\rm R}}{2k_{\rm{F}}}\left( f^{(1)}_+\frac{u^{(-1)}_2+u^{(1)}_2}{2}+f^{(1)}_-\frac{u^{(-1)}_3-u^{(1)}_3}{2i} \right) \right].
\eea
Here, in addition to Eq. (\ref{eq:fs}) we have defined
\bea\label{eq:hs}
&h^{(m)}=1+H_m,~~~h^{(m)}_{+}=1+\frac{H_{m-1}+H_{m+1}}{2},~~~h^{(m)}_{-}=\frac{H_{m-1}-H_{m+1}}{2}.&
\eea
Like in the 2DEG case, renormalization of the couplings (via $f^{(m)}_\pm$, $f^{(m)}$, $h^{(m)}_\pm$ and $h^{(m)}$ factors) leads to splitting of resonances in the in-plane and out-of-plane modes. In addition, electron-electron interaction also couples the spin-chiral and spin-valley degrees of freedom leading to an additional splitting of resonance frequencies. Solving for the eigenvalues of the system in the $m=1$ channel [Eqs. \eqref{eq:EomGr-intE} and \eqref{eq:EomGr-intB} in \cref{AppGRa}], we get the following resonant frequencies \cite{kumar2021}:
\bea\label{eq:reosnancefreq}
\Omega^2_\pm &=& \Gamma^2\pm\Omega_0^2,\nn\\
\text{where}~\Gamma^2&\equiv&\left(\dfrac{f^{(1)}f_+^{(1)}+h^{(1)}h_+^{(1)}}{2}\right)\lambda_{\rm R}^2+(f_-^{(1)}h_-^{(1)}+f_+^{(1)}h_+^{(1)})\lambda_{\rm Z}^2,\;\text{and}\nn\\
\Omega_0^2&\equiv&\left[\left(\dfrac{f^{(1)}f_+^{(1)}-h^{(1)}h_+^{(1)}}{2}\right)^2\lambda_{\rm R}^4+(f_+^{(1)}h_-^{(1)}+f_-^{(1)}h_+^{(1)})^2\lambda_{\rm Z}^4+(f^{(1)}f_-^{(1)}+h^{(1)}h_-^{(1)})(f_+^{(1)}h_-^{(1)}+f_-^{(1)}h_+^{(1)})\lambda_{\rm R}^2\lambda_{\rm Z}^2\right]^{1/2}.\nn\\
\eea

The response to the electric-field driving yields the following current density
\bea\label{eq:jGre}
\bf J&=&\underbrace{\frac{i\sigma_0}{\tau}\left[\frac{f^{S,(1)}}{\bar\Omega }+\frac{1}{\bar\Omega}\left(\frac{\bar\Omega}{2\mu}\right)^2R^{\rm Di}_{\rm JE}(\Omega)\right]}_{\sigma(\Omega)}{\bf E}_0,\nn\\
\text{where}~R^{\rm Di}_{\rm JE}(\Omega)&=&\dfrac{W^{\rm JE}_+}{(\bar\Omega^2-\Omega_+^2)}-\dfrac{W^{\rm JE}_-}{(\bar\Omega^2-\Omega_-^2)},\nn\\
W^{\rm JE}_\pm &=&\dfrac{f_+^{(1)}\lambda_{\rm R}^2}{2\Omega_0^2}[\Omega_\pm^2-\Omega_{\rm JE}^2],~~\Omega_{\rm JE}^2=h^{(1)}h_+^{(1)}\lambda_{\rm R}^2+\frac{[(f_+^{(1)})^2-(f_-^{(1)})^2]h_+}{f_+^{(1)}}\lambda_{\rm Z}^2,
\eea
and magnetization
\bea\label{eq:maggr2}
{\bf M}&=&\underbrace{\dfrac{-i\sigma^{\rm ME}_0}{\bar\Omega\tau}\left(\dfrac{\bar\Omega}{\lambda_{\rm R}}\right)^2R^{\rm Di}_{\rm ME}(\Omega)}_{\sigma^{\rm ME}(\Omega)} {\bf E}_0\times \hat{z};\nn\\
M_z&=&0,\nn\\
\text{where}~R^{\rm Di}_{\rm ME}(\Omega)&=&\dfrac{W^{\rm ME}_+}{(\bar\Omega^2-\Omega_+^2)}-\dfrac{W^{\rm ME}_-}{(\bar\Omega^2-\Omega_-^2)},\nn\\
W^{\rm ME}_\pm&=&\dfrac{\lambda_{\rm R}^2}{2\Omega_0^2}[\Omega_\pm^2-\Omega_{\rm ME}^2],\\
\Omega^2_{\rm ME}&=&h^{(1)} h_+^{(1)}\lambda_{\rm R}^2+(f_+^{(1)}-f_-^{(1)})(h_+^{(1)}-h_-^{(1)})\lambda_{\rm Z}^2.
\eea

\begin{figure*}[htp]
\centering
\includegraphics[width=0.35\linewidth]{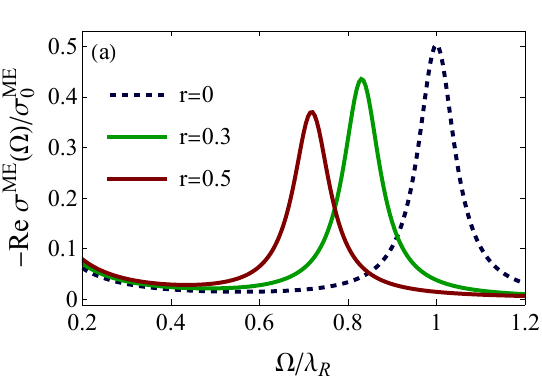}~
\includegraphics[width=0.35\linewidth]{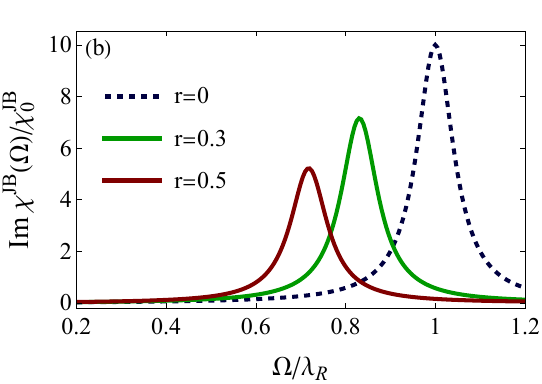}~
\\
\includegraphics[width=0.35\linewidth]{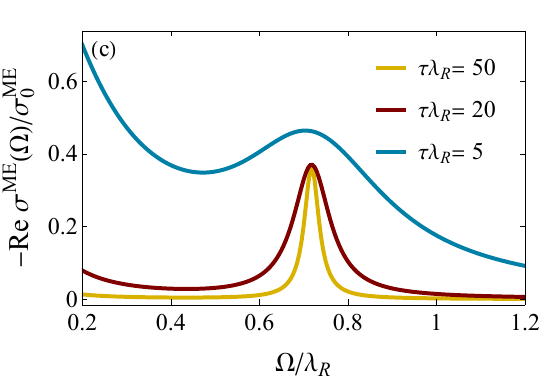}~
\includegraphics[width=0.35\linewidth]{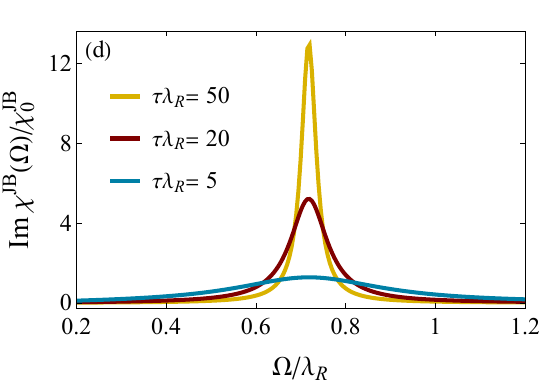}~
\\
\includegraphics[width=0.35\linewidth]{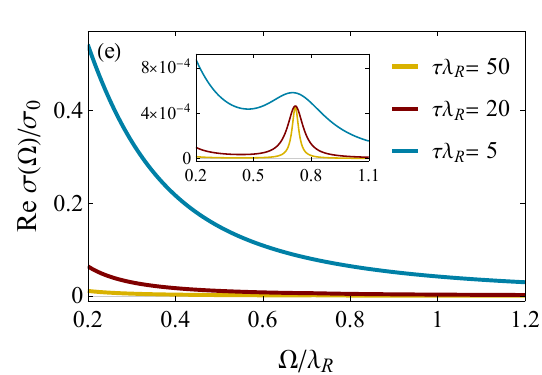}~
\includegraphics[width=0.35\linewidth]{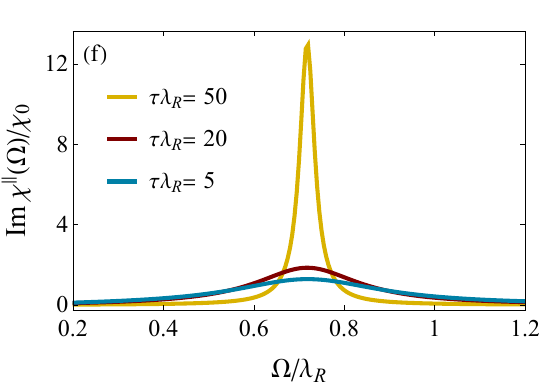}~
\caption{\textbf{Dynamical response of a single-valley system with Rashba spin-orbit coupling.} Panels (a) and (b) display the evolution of the cross-response functions, Re$\sigma^{\rm ME}(\Omega)$ and Im$\chi^{\rm JB}(\Omega)$, respectively, with a dimensionless interaction parameter $r$, defined by $F^A_0=-r,F^A_1=-r/2,F^A_2=-r/4,F_1^S=r/6$. The values of other parameters are chosen as $\tau\lambda_{\rm R}=20$ and $\mu=5\lambda_{\rm R}$. Panels (c) and (d) depict the behavior of the same response functions as in (a) and (b), respectively, with increasing damping, while the interaction parameter is  fixed at $r=0.5$. Panels (e) and (f) depict the direct response functions, Re$\sigma(\Omega)$ and Im$\chi^{||}(\Omega)$, respectively,
for the same set of parameters as in panels (c) and (d). The inset in panel (e) shows the resonance in the conductivity with the Drude background subtracted.}\label{fig:2DEG}
\end{figure*}

\begin{figure*}[htp]
\centering
\includegraphics[width=0.35\linewidth]{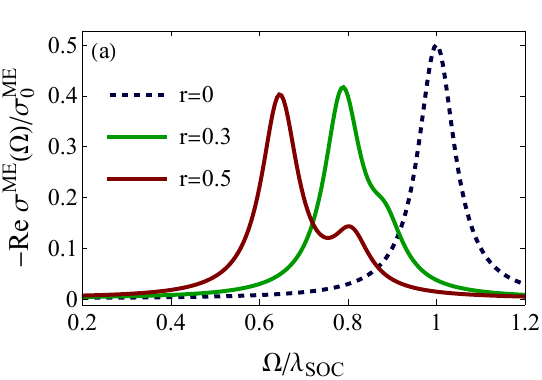}~
\includegraphics[width=0.35\linewidth]{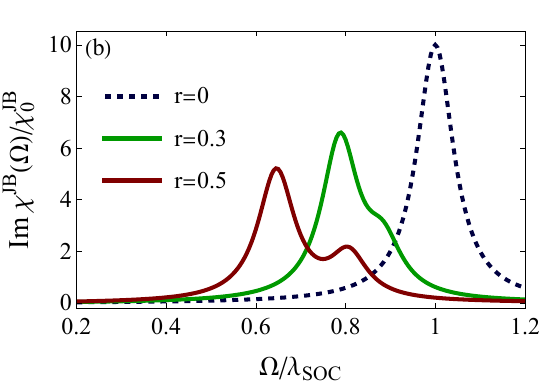}~
\\
\includegraphics[width=0.35\linewidth]{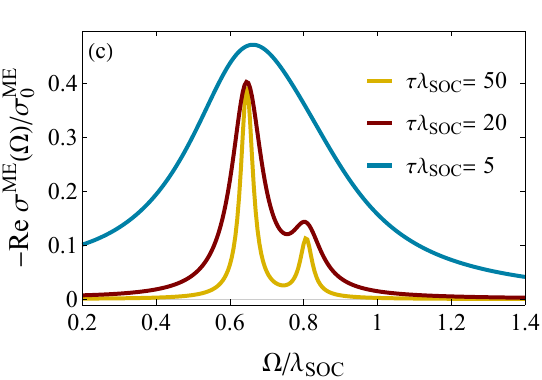}~
\includegraphics[width=0.35\linewidth]{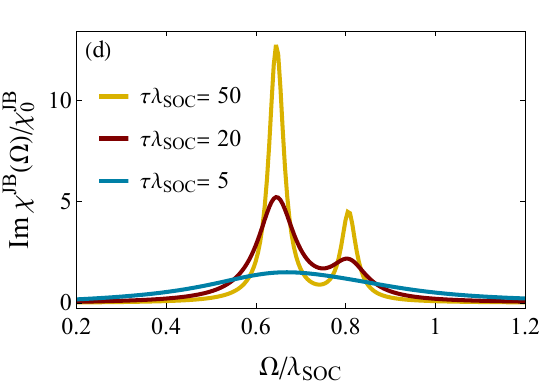}~
\\
\includegraphics[width=0.35\linewidth]{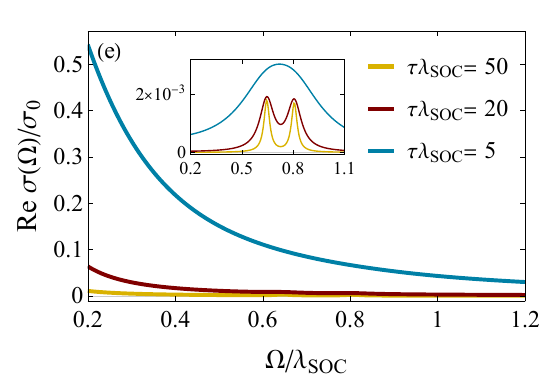}~
\includegraphics[width=0.35\linewidth]{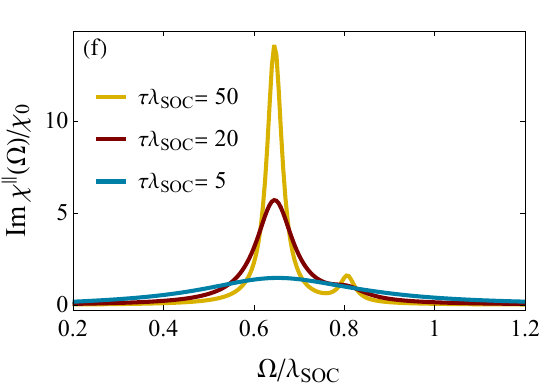}~
\caption{\textbf{Dynamical response of a two-valley system with Rashba and valley-Zeeman spin-orbit couplings.} Panels (a) and (b) display the evolution of the cross-response functions, Re$\sigma^{\rm ME}(\Omega)$ and Im$\chi^{\rm JB}(\Omega)$, respectively, for increasing values of the interaction parameter $r$, such that $F^A_0=-r,F^A_1=-r/2,F^A_2=-r/4,F_1^S=r/6$, and for a fixed value of $H_{0,1,2}=0.9F^A_{0,1,2}$. The values of other parameters are chosen as $\tau\lambda_{\rm SOC}=20,\mu=5\lambda_{\rm SOC}$, $\lambda_{\rm SOC}=\sqrt{\lambda^2_{\rm R}+\lambda^2_{\rm Z}}$, and $\lambda_{\rm Z}=0.5\lambda_{\rm R}$. Note that the splitting of the resonance peak increases with interaction. Panels (c) and (d) depict the behavior of the same response functions as in (a) and (b), respectively, with increasing damping, at fixed interaction parameter $r=0.5$. Panels (e) and (f) depict the direct response functions, Re$\sigma(\Omega)$ and Im$\chi^{||}(\Omega)$, respectively, for the same set of parameters as in panels (c) and (d). The inset in panel (e) shows the resonance in the conductivity with the Drude background subtracted.}\label{fig:Di}
\end{figure*}

For $B$-field driving, the current density is given by
\bea\label{eq:jGrb}
\bf J&=&\underbrace{-\chi_0^{\rm JB}R^{\rm Di}_{\rm JB}(\Omega)}_{\chi^{\rm JB}(\Omega)}{\bf B}_0\times\hat z,
\label{JB}
\\
\text{where}~R^{\rm Di}_{\rm JB}(\Omega)&=&\dfrac{W^{\rm JB}_+}{\bar\Omega^2-\Omega_+^2}-\dfrac{W^{\rm JB}_-}{\bar\Omega^2-\Omega_-^2},\nn\\
W^{\rm JB}_\pm&=&\dfrac{f^{(1)}f_+^{(1)}\lambda_{\rm R}^2+(f_+^{(1)}+f_-^{(1)})(h_+^{(1)}+h_-^{(1)})\lambda_{\rm Z}^2}{2\Omega_0^2}(\Omega_\pm^2-\Omega^2_{\rm JB}),\\
\Omega_{\rm JB}^2&\equiv&\dfrac{[f_+^{(1)}h^{(1)}\lambda_{\rm R}^2+(f_+^{(1)}-f_-^{(1)})(f_+^{(1)}+f_-^{(1)})\lambda_{\rm Z}^2][f^{(1)}h_+^{(1)}\lambda_{\rm R}^2+(h_+^{(1)}-h_-^{(1)})(h_+^{(1)}+h_-^{(1)})\lambda_{\rm Z}^2]}{f^{(1)}f_+^{(1)}\lambda_{\rm R}^2+(f_+^{(1)}+f_-^{(1)})(h_+^{(1)}+h_-^{(1)})\lambda_{\rm Z}^2},
\eea
while the magnetization reads
\bea\label{eq:MGb}
\bf M&=&\underbrace{-\frac{\chi_0}2R^{\parallel \rm Di}_{\rm MB}(\Omega)}_{\chi^\parallel(\Omega)}{\bf B}_0,\;
M_z=\underbrace{-\chi_0R^{\perp \rm Di}_{\rm MB}(\Omega)}_{\chi^\perp(\Omega)}B_z,\nn\\
\text{where}~R^{\parallel \rm Di}_{\rm MB}(\Omega)&=&\dfrac{W^{\rm MB}_+}{\bar\Omega^2-\Omega_+^2}-\dfrac{W^{\rm MB}_-}{\bar\Omega^2-\Omega_-^2},\nn\\
W^{\rm MB}_{\pm}&=&\dfrac{f^{(1)}\lambda_{\rm R}^2+2(h_-^{(1)}+h_+^{(1)})\lambda_{\rm Z}^2}{2 \Omega_0^2}[\Omega_\pm^2-\Omega^2_{\rm MB}],\\
\Omega_{\rm MB}^2&\equiv&\dfrac{[h^{(1)}\lambda_{\rm R}^2+2(f_+^{(1)}-f_-^{(1)})\lambda_{\rm Z}^2][f^{(1)}h_+^{(1)}\lambda_{\rm R}^2+(h_+^{(1)}-h_-^{(1)})(h_+^{(1)}+h_-^{(1)})\lambda_{\rm Z}^2]}{f^{(1)}\lambda_{\rm R}^2+2(h_+^{(1)}+h_-^{(1)})\lambda_{\rm Z}^2},\nn\\
R^{\perp \rm Di}_{\rm MB}(\Omega)&=&\dfrac{f_+^{(0)}\lambda_{\rm R}^2}{\bar\Omega^2-f^{(0)}f_+^{(0)}\lambda_{\rm R}^2-f_+^{(0)}h_+^{(0)}\lambda_{\rm Z}^2}.\label{eq:MGb2}
\eea
The three resonant frequencies are $\Omega_+, \Omega_-$, and $\Omega_z=\sqrt{f^{(0)}f^{(0)}_+\lambda_{\rm R}^2+f^{(0)}_+h^{(0)}_+\lambda^2_{\rm Z}}$. The first two correspond to the in-plane CSMs, which are split into two due to the coupling between the spin-chiral and spin-valley degrees of freedom. Both the $\Omega_+$ and $\Omega_-$ modes show up in cross-response. The $\Omega_z$ mode corresponds to oscillations of $M_z$ and does not appear in cross-response. Note that there is one more out-of-plane mode, identified in Ref.~\cite{kumar2021}, which does not couple to macroscopic electric and magnetic fields, but can be induced by valley spin population imbalance. In the limit of $\lambda_{\rm Z}\rightarrow 0$, this mode corresponds to out-of-plane oscillations of the magnetization difference between the two valleys.

\subsection{Lineshapes of the cross-response resonances}
\label{sec:analysis}
\begin{figure*}[htp]
\centering
\includegraphics[width=0.85\linewidth]{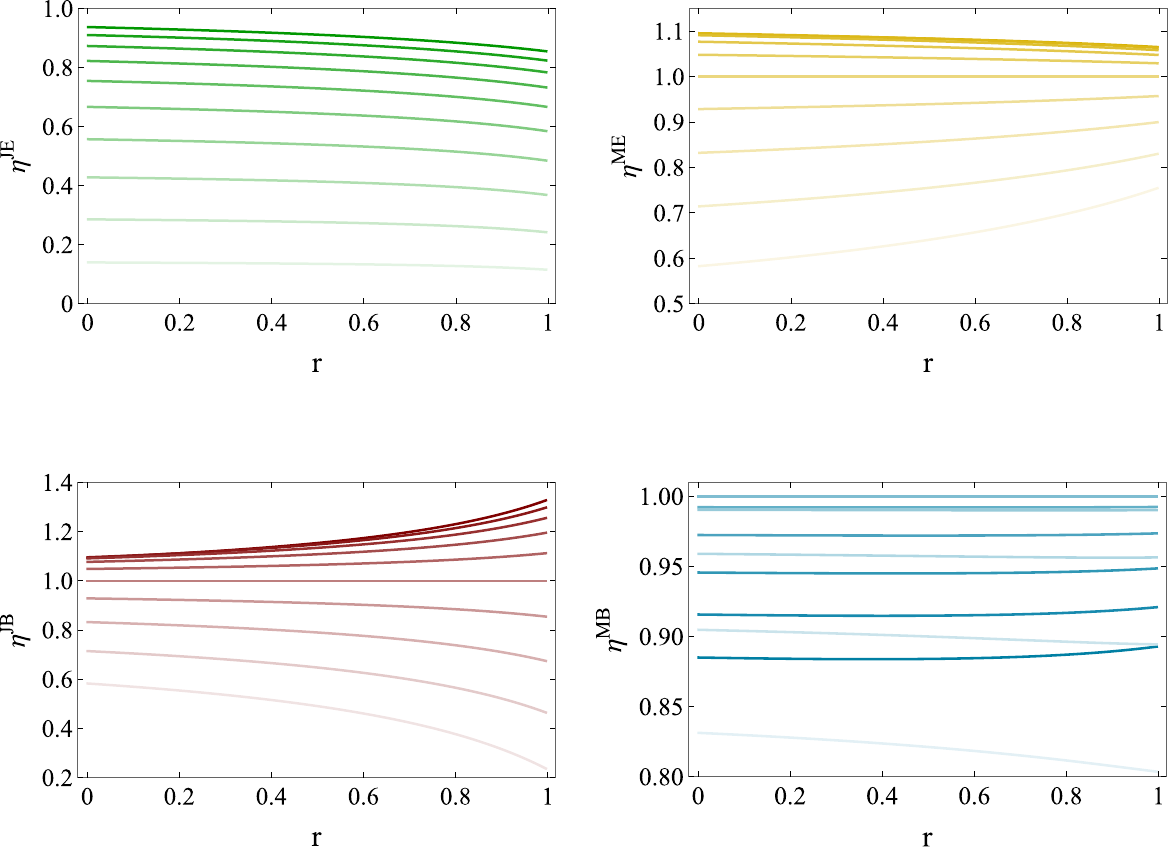}
\caption{\textbf{Ratio of the spectral weights of the interaction- split resonances for a two-valley system.} The variation of the ratio $\eta^{aa'}$ with the parameter $r$. The shades from dark to light in each case correspond to $u\equiv H_{m}
/F^{A}_{m}\in\{0,0.1,...,0.9\}$.\label{fig:Weights}}
\end{figure*}

In this section, we analyze the lineshapes of resonances in the presence of \emph{eei}, which are markedly different for single- and two-valley systems due to \emph{eei}. Figures~\ref{fig:2DEG}(a) and \ref{fig:2DEG}(b) show the cross-response functions, Re$\sigma^{\rm ME}(\Omega)$ and Im$\chi^{\rm JB}(\Omega)$, respectively, for a single-valley system as a function of the interaction parameters $F^A_{\{0,1,2\}}$, as specified in the figure caption. We see that the interaction leads to a downward shift of the resonance frequency and a reduction of its amplitude. Panels c) and (d) of the same figure show how the resonances in the cross-response functions evolve with damping, while panels (e) and (f) display the corresponding behavior for the direct response functions, Re$\sigma(\Omega)$ and Im$\chi^{||}(\Omega)$. We observe that the resonant peaks in Im$\chi^{\rm JB}(\Omega)$ and  Im$\chi^{||}(\Omega)$ behave with damping in the usual way, becoming both broader and lower as damping increases. In contrast, while the peaks in Re$\sigma^{\rm ME}(\Omega)$ and Re$\sigma(\Omega)$ broaden with increasing damping, their amplitudes remain almost independent of damping for  $\tau\lambda_{\rm R}\gg 1$, and actually increases with damping for $\tau\lambda_{\rm R}\gtrsim 1$. This unusual behavior arises from the interplay between the frequency dependence of the resonant and non-resonant factors in Eq.~\eqref{eq:m2DEGe}, and also from the fact that both conductivities are normalized to the \emph{dc} values, which by themselves are proportional to $\tau$. 

Figure \ref{fig:Di} displays both direct and cross-response functions for a two-valley system. The main difference compared to the single-valley case is that each resonance is now split into two. This is due to  \emph{eei} which couple excitations in the spin and spin-valley sectors. As damping increases,  the resonances in the two-valley system behave similarly to those in a single-valley system.

Next, we observe that, in the two-valley system, the spectral weight is distributed very unevenly between interaction-split resonances in $\sigma^{\rm ME},\chi^{\rm JB},$ and $\chi$, but more evenly for $\sigma$. To elucidate this point, we analyze the ratio of the spectral weights as a function of the interaction parameters. According to Eqs.~(\ref{eq:jGre})-(\ref{eq:MGb2}), the spectral weight of a resonance at frequency $\Omega_\pm$ is given by $\pm W^{aa'}_{\pm}/2\Omega_\pm$, where $a=J,M$ and $a'=E,B$. To compare the spectral weights of the modes in various response functions, we introduce a dimensionless 
quantity--the difference in spectral weights of the lower and higher energy modes divided by their sum:
\bea\eta^{aa'}\equiv \frac{(-W^{aa'}_-/2\Omega_-)-W^{aa'}_+/2\Omega_+}{(-W^{aa'}_-/2\Omega_-)+W^{aa'}_+/2\Omega_+}.\eea Using Eqs. (\ref{eq:jGre})-(\ref{eq:MGb2}), we arrive at
\bea\label{eq:ratioof}
\eta^{aa'}&=&\left(\frac{\Omega_++\Omega_-}{\Omega_+-\Omega_-}\right)\frac{\Omega^2_{aa'}-\Omega_+\Omega_-}{\Omega^2_{aa'}+\Omega_+\Omega_-},
\eea
where $\Omega_{aa'}$ are introduced in Eqs.~(\ref{eq:jGre})-(\ref{eq:MGb2}). If the ratio $\eta^{aa'}$ is close to one (minus one), the spectral weight is predominantly carried by the lower (higher) energy mode, while if it is close to zero, the two modes carry comparable spectral weights. We see that, for fixed resonance frequencies $\Omega_\pm$, the distribution of spectral weights across different responses $aa'$ is controlled by the ratio $\Omega^2_{aa'}/\Omega_+\Omega_-$. Variations in $\Omega_{aa'}$ across different responses lead to distinct spectral weight allocations between the modes. Figure \ref{fig:Weights} displays the ratio $\eta^{aa'}$ as a function of the interaction parameter $r$, defined in the caption to Fig.~\ref{fig:Di}, for a range of  inter-valley interactions in the spin sector parameterized by $H_m=uF^A_m$, with $u$ being the same for all $m$. We observe that the values $\eta^{aa'}$ are clustered in the interval $(\sim 0.5,1)$ for the ME and JB responses, and in the interval $(\sim 0.9,1)$ for the MB response, which means that spectral weight is concentrated mostly in the lower-energy mode, for a wide range of both $r$ and $u$. In contrast, the JE response exhibits a large variation the spectral weights ratio, which is close to 1 for smaller values of $u$ but quickly approaches zero for larger values. The parameters chosen in Fig.~\ref{fig:Di} happen to correspond to a strong inter-valley interaction, which explains why the two peaks in the conductivity (panel (e) in Fig.~\ref{fig:Di}) are of comparable strength.

\section{Connection to the experiment
}\label{Sec:Exp}
In this section, we discuss the prospects of detecting the resonances, described in the previous sections, in the experiment. In what follows, we focus on the two-valley systems, where neither direct nor cross-response resonances due to CSMs have been detected yet, and emphasize the advantages of measuring the cross-responses, and propose potential applications of the resonant cross-responses in spintronics.

\subsection{Detecting resonances in cross-response}
As seen from \cref{fig:1}, the $E$-field driven responses, both direct $J^E$ and cross $M^E$, are stronger that the magnetically driven counterparts (the separation between the green and red lines in the figure panels). While the finiteness of the cross-response is due to SOC, the dominance of the electrically driven response is due to the stronger magnitude of the $E$-field. However, there is an additional boost (of about another order of magnitude) from coupling to the CSM of the system. This is evident from peak in  \cref{fig:1}(b) around $\Omega\sim\lambda_{\rm SOC}$. This boost is pronounced much stronger in the cross-response sector ($M^E$), via the resonant Edelstein effect, than in the direct sector ($J^E$). This is so because the resonance in $J^E$ is typically masked by the Drude tail, unless SOC is very strong, and one needs to find a way to subtract the Drude contribution from the data (we remind the reader that \cref{fig:1} is for non-interacting electrons and hence there is only peak). Additionally, in a 2DEG where there can be Rashba and Dresselhaus types of SOC, we see from the tensor structure of the cross-response function $\sigma^{\rm ME}$ (\cref{eq:TensorStructures}) that different components are proportional to different types of SOC. These observations suggest that i) the 
CSMs resonances would be better resolved in cross-response measurements, and ii) detecting the resonance in different components of the cross-response tensor would allow us to identify the dominant type of SOC in the system (in 2DEGs). In 
direct response, the effect of different types of SOC are not distinguishable, and hence a detailed modeling is necessary to extract Rashba and/or Dresselhaus coupling constants from the data \cite{Baboux2013,Baboux2015,Perez2016,maiti2017}. In contrast, cross-response offers a more straightforward
approach to identify the dominant type of SOC. To estimate the size of the cross-response effect, note that with pulsed $E$-fields on the order of a few MV/cm ~\cite{kim:2014, huber:2019, yutong:2021}, the peak magnetization near CSRs from Eq. \eqref{eq:jGre12} is expected to be $\sim (\lambda_{\rm SOC}\tau/250)\mu_B$ per square nm of the sample.  (For graphene on TMD,  we have $\lambda_{\rm SOC}\tau\sim 5$~\cite{morpurgo:prx}). Since $\lambda_{\rm SOC}$ is in the meV range, the resonance frequencies lie in the THz range. Correspondingly, the wavelength of THz radiation is in mm range and, to avoid finite size effects, one would need samples of larger dimensions. In this regard, it is worth noting that it is now possible to grow large graphene samples which are suitable for THz spectroscopy\cite{Mead2025}.

\begin{figure*}[htp]
\centering
\includegraphics[width=0.45\linewidth]{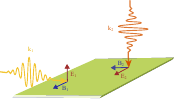}
\caption{\textbf{Probing the resonant inverse Edelstein (spin-galvanic) effect}: When an electromagnetic pulse (labeled 1) is incident at a grazing angle to the 2D plane, such that the $E$-field is polarized perpendicular to the plane, the in-plane Drude response 
is absent. The response would be primarily due to the in-plane field $B_1$. In case of normal incidence (pulse 2), the current induced by the $B$-field $B_2$ would be masked by the Drude part of the current 
induced by the $E$-field $E_2$.}\label{Fig:RIEF}
\end{figure*}

\subsection{Detecting CSM via resonant inverse-Edelstein effect} 
In this effect, an oscillatory $B$-field drives an electric current. As one can see from Eq.~\eqref{JB}, the induced current is perpendicular to the direction of the $B$-field.  For a normal incidence of an electromagnetic pulse, both $E$- and $B$-fields are  in the 2DEG plane. Therefore, the magnetically-driven current is along the $E$-field of the pulse, and thus also along the electrically-driven current. As we already said, the latter contains a large Drude component, which masks both the EDSR part of $J^E$ and $J^B$. However, if the electromagnetic wave is incident at a grazing angle, with the $E$-field pointing out of the plane, as shown in  \cref{Fig:RIEF}, the Drude response 
is absent, thus allowing full access to the resonance due to the inverse Edelstein effect. 
\begin{figure}[htp]
\centering
\includegraphics[width=0.9\linewidth]{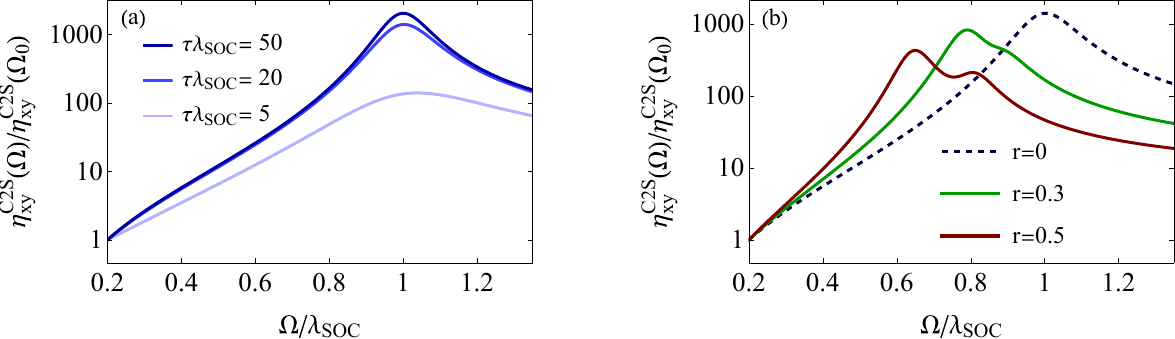}
\caption{\textbf{Resonant charge-to-spin (C2S) conversion}. (a) Enhancement in the C2S conversion merit tensor in a two-valley system near the CSM resonance frequency in the non-interacting limit. $\Omega_0=0.2\lambda_{\rm SOC}$ is a reference frequency, chosen sufficiently far away from the resonance. (b) The same as (a) but with electron-electron interactions included via the parameter $r$, as defined in the caption to \cref{fig:Di}. Here, $\tau\lambda_{\rm SOC}=20$, and the remaining parameters are the same as specified in \cref{fig:Di}.}\label{fig:2}
\end{figure}
\subsection{Applications to spintronics}
\paragraph*{Resonant charge-to-spin conversion:} The efficiency of charge-to-spin (C2S) conversion by a spintronic device, operating in the \emph{dc} regime, is commonly quantified by an intrinsic figure-of-merit, $\eta^{\rm C2S}\equiv e\sigma^{\rm ME}(0)/\mu_B\sigma(0)$, with units the inverse velocity~\cite{Offidani2017}. The highest values of $v_{\rm F}\eta\sim 0.1-3$~\cite{Ghiasi2019,Lijun2020,Monaco2021} are achieved when $\mu$ is tuned to the bottom of the Rashba-split bands, so that only the lowest subband is occupied. This can be a challenging regime to enter as the doping has to be fine-tuned to access this region. For a more common case of larger $\mu$, where both subbands are occupied $v_{\rm F}\eta$ is significantly smaller~\cite{Offidani2017}. Here, we extend the quantification of C2S conversion to the dynamical regime, by defining a figure-of-merit tensor,
$\eta^{\rm C2S}_{\alpha\beta}(\Omega)\equiv|e\sigma^{\rm ME}_{\alpha\beta}(\Omega)/\mu_B\sigma_{\alpha\beta}(\Omega)|$.  In Fig. \ref{fig:2}a,  we demonstrate that operating near a CSM resonance frequency leads to a substantial enhancement--by a factor of$\sim 10^{2}-10^3$--in C2S conversion merit tensor, as compared to its value away from the resonance. While Fig. \ref{fig:2}a is plotted for a Dirac system, the result is similar for a 2DEG. What is noteworthy for the 2DEG case is that the enhancement happens in the large-$\mu$ limit, where the conversion is known to be suppressed in the \emph{dc} regime. 

Figure \ref{fig:2}b shows the effect of \textit{eei} on $\eta^{\rm C2S}_{xy}(\Omega)$. We see two peaks in $\eta^{\rm C2S}_{xy}(\Omega)$ which essentially track the two CSMs which split with increasing interaction parameter $r$ and also shift to lower energies. The resolution of the peaks is dependent on disorder. In Fig. \ref{fig:2}b we used $\tau\lambda_{\rm SOC}=20$.

\begin{figure*}[htp]
\centering
\includegraphics[width=0.9\linewidth]{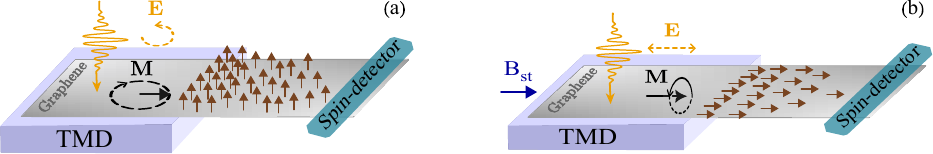}
\caption{\textbf{Architectures for spin-pumping.} 
Panels a) and (b) illustrate proposed architectures for electrically driven, ultrafast THz spin-pumping in a graphene layer, part of which is attached to a TMD substrate. The latter induces SOC in the adjacent layer. In panel (a), a circularly polarized pulse injects spins, polarized in the out-of-plane direction, into a SOC-free region. In panel (b), a linearly polarized pulse is applied in combination with an static magnetic field, $B_{\rm st}$, which controls the polarization of 
injected spins. The direction of spin polarization in the SOC-free part of graphene is shown by little arrows. A spin detector, e.g. a Pt strip, registers the spin current via the inverse spin Hall effect. The density of arrows conveys diffusive spreading of the spins.}\label{fig:spinpump}
\end{figure*}

\paragraph*{Architecture for ultra-fast spin-pumping.} Precessing magnetization of a magnet pumps spins to an adjacent non-magnetic layer (without pumping charge) ~\cite{Tserkovnyak2002,Costache2006}, producing a flow of spins $\propto g_s\mathbf M\times d \mathbf M/dt$, which are polarized along the precession axis\footnote{There can be an additional channel $\propto dM_a/dt$ if such contacts can be designed ~\cite{Tserkovnyak2002}.}. Here, $g_s$ is the spin-mixing conductance
which is controlled by the properties of an interface between the magnetic and non-magnetic layers. Conventionally, spin-pumping is achieved by exciting a ferromagnetic resonance (FMR), which lies typically in the GHz range. Recently, ultrafast spin-pumping using antiferromagnetic resonance (aFMR) in the sub-THz regime has also been demonstrated ~\cite{Vaidya2020,Li2020}. 
Given that the SO-induced spin-splitting in graphene on TMD is in the THz range ~\cite{morpurgo:prx, bockrath:2019}, we propose a different architecture for ultrafast spin-pumping, based on the resonant Edelstein effect. The proposed setup is shown in Fig.~\ref{fig:spinpump}, panels (a) and (b). Panel (a) depicts an all-electric platform, in which a circularly polarized THz pulse is incident on a region of graphene with both Rashba and VZ types of SOC induced by a TMD substrate. Via the resonant Edelstein effect, the circularly polarized $E$-field of the pulse induces a circularly polarized magnetization, which induces electron spins  to precess about the normal to the plane. Since the SOC-free side of graphene is not spin-polarized, precessing spins flow out of the region in contact with TMD towards a detector, which registers the spin-current via the inverse spin-Hall effect. The TMD-free side may need to be masked, to be protected from the THz pulse. The SOC-free region need not be suspended, but rather mounted on a SOC-inert substrate. Since the precession axis in the SOC-coupled region is out-of-plane, the direction of pumped spins in the SOC-free region is also out-of-plane.

One can also inject in-plane polarized spins using a linearly polarized pulse in combination with a static in-plane $B$-field ($\mathbf B_{\rm st}$), as depicted in Fig.~\ref{fig:spinpump}(b). While the induced magnetization is linearly polarized in the absence of $\mathbf B_{\rm st}$, the latter introduces elliptic polarization with its axis along $\mathbf B_{\rm st}$. The ellipticity is proportional to $|\mathbf B_{\rm st}|$~\cite{maiti2016,kumar2017}. Since the precession axis of magnetization is fixed by the direction of $\mathbf B_{\rm st}$, the orientation of pumped spins is controlled by the orientation of $\mathbf B_{\rm st}$ as well.

In the above proposals, graphene was chosen due to its long spin relaxation length ~\cite{Ingla2015,Drogeler2016} (in the absence of a TMD substrate), as well as the demonstrated feasibility of on-chip spin injection into graphene using ferromagnetic contacts\cite{Indolese2018}. 
According to Ref.~~\cite{morpurgo:prx}, $\lambda_{\rm R}\tau\sim 5$ in graphene on a TMD substrate, which makes the resonance reasonably sharp.

\section{Conclusions}\label{Sec:Conclusion}
While the \emph{dc} Edelstein effect and its inverse--the spin-galvanic effect--along with their roles in spin-charge conversion have been studied in the past, this  work focuses on their resonant counterparts in the presence of 
electron-electron interactions (\emph{eei}). The resonance arises from the chiral-spin modes (CSMs)--collective excitations of a Fermi liquid with spin-orbit coupling (SOC), induced by broken inversion symmetry. We analyzed the similarities and differences of the above effects in single-valley (2D electron gas) versus two-valley (monolayer graphene on TMD) systems.

In particular, we demonstrated that, in the non-interacting limit, the both the resonant Edelstein effect and its inverse occur in the same way in single and two-valley systems with Rashba type of SOC. Out of three CSMs, only two in-plane modes give rise to resonances in cross-response. We also showed that electrically driven responses (both direct or cross) are stronger than magnetically driven ones. While a resonance in direct response, which occurs in the optical conductivity $\sigma(\Omega)$ via  the EDSR effect, is not well resolved due to a strong Drude background, a resonance in the cross-response function, i.e., the Edelstein (or magnetoelectric) conductivity $\sigma^{\rm ME}(\Omega)$, is very prominent. For a single valley system, a resonance appears in different components of the response tensor, depending on where the dominant SOC of the Rashba or Dresselhaus type is. While a two-valley system also has the valley-Zeeman (VZ) SOC which affects the resonance frequency, CSMs occur in cross-response only due to Rashba SOC. We also suggested that the resonant inverse-Edelstein effect, driven by an oscillatory $B$-field, is best observed at grazing incidence of an electromagnetic pulse.

Electron-electron interaction (\textit{eei}) renormalize the resonance frequencies and alter the line shapes both in single- and two-valley systems. In the two-valley system, \textit{eei} splits the resonance  into two peaks, corresponding to coupled oscillations in spin-chiral and spin-valley sectors. We showed that, for a wide range of interaction parameters, the spectral weight is distributed unevenly between the two peaks such that the lower energy mode carries a larger spectral weight.

In view of potential applications in spintronics, we showed that a commonly used figure of merit, $\eta(\Omega)\propto\sigma^{\rm ME}(\Omega)/\sigma(\Omega)$, which quantifies the efficiency of charge-to-spin conversion, is enhanced near a resonance by several orders magnitude, and that too in the regime of 
large number densities, where the \emph{dc} counterpart of $\eta$ is suppressed. We also suggested an architecture for pumping resonantly excited spins and a way to control the direction of the excited spins, using either circularly polarized light or linearly polarized light in combination with an in-plane static magnetic field. Both these properties are of prime interest in spintronics.

To conclude, one can state that, in general, the response of a conductor with SOC to oscillatory $E$- and $B$-fields can be described by the following relations:
\bea\label{eq:conc}
J_\alpha &=\underbrace{\sigma_{\alpha\beta}(\Omega)}_\text{Drude+EDSR+Hall}E_\beta &+ \underbrace{\chi^{\rm JB}_{\alpha\beta}(\Omega)}_\text{Resonant inverse-Edelstein} B_\beta,\nn\\
M_\alpha&=\underbrace{\sigma^{\rm ME}_{\alpha\beta}(\Omega)}_\text{Resonant Edelstein} E_\beta &+~~~~~~~~~~\underbrace{\chi_{\alpha\beta}(\Omega)}_\text{Chiral-spin resonance}B_\beta.
\eea
The underbraces below the response functions denote the effects that they host. The tensor structure of the cross-response functions depends on the type of SOC [cf. Eq.~(\ref{eq:TensorStructures})], which may be used to identify various types of SOC present in a given material. With respect to the polarization of the oscillatory fields, we showed that the in-plane $E$ and $B$-fields induce both direct and cross-responses, while the out-of-plane $B$-field induces only direct response. We also point out that cross-responses in insulators has recently been considered in Ref.~\cite{Duff2022}. Having understood the electromagnetic response from the chiral-spin modes, one could also explore their role in the Chiral-Induced Spin Selectivity phenomena\cite{Bloom2024} that has been gaining traction in recent years.

\section{Rashba: Hamiltonian and Person}

As is the case for many papers in the field of spintronics, the key physical ideas discussed in the present work can be traced back to the seminal contributions of Emmanuel Iosifovich Rashba (E.I.R.): the early works by Rashba and Sheka \cite{Rashba_Sheka}, and by Bychkov and Rashba \cite{Bychkov1984}, as well as a closely related paper by E.I.R.’s former student, Victor Edelstein \cite{EDELSTEIN1990}.

One of us (D.L.M.) had the privilege of knowing E.I.R. since his graduate school days. Although his own research interests at the time were far from spintronics, the seeds of appreciation for the beauty of spin-orbit coupling were planted through E.I.R.'s talks, interactions with his disciples, and personal discussions. These seeds came to fruition much later, during a direct collaboration with E.I.R. on developing a Fermi-liquid theory in the presence of spin-orbit coupling \cite{ashrafi:2013}.

That project was ambitious and took more than a year to complete. The collaboration with E.I.R. was both immensely rewarding and intellectually rigorous. Although the Fermi-liquid theory was not part of his daily toolkit, E.I.R. left no stone unturned and insisted on absolute clarity in every statement. We went through more than 30 versions of the manuscript before it met his high standards.

Despite his advanced age and ongoing health issues at the time, his energy and dedication were remarkable. After nearly every phone conversation, D.L.M. would receive a neatly typed note summarizing the points discussed. On a lighter note, when E.I.R. happened to call D.L.M. after 10 p.m., he would apologize—not for the late hour, but for \emph{interrupting} his work. 

If one were to search for a single word to characterize E.I.R.—both as a scientist and as a person—the most fitting choice would be \emph{correct}. Correct in his theoretical predictions, correct in his interpretations of experiments, and correct and respectful in his interactions with colleagues and co-authors, from graduate students to distinguished professors alike.

\section{Acknowledgements} 
We thank P. Armitage, C. R. Bowers, and X.-X. Zhang for stimulating discussions. M.S. and S.M. were funded by the Natural Sciences and Engineering Research Council of Canada (NSERC) Grant No. RGPIN-2019-05486. A.K. acknowledges support from Canada First Research Excellence Fund and by the Natural Sciences and Engineering Research Council of Canada (NSERC) under Grant No. RGPIN-2019-05312. DLM was supported by the US National Science Foundation via grant DMR-2224000. 

\appendix

\section{
Equations of motion a single-valley system}\label{App2DEG}
Taking the Fourier transform of the kinetic equation [\cref{eq:EdriveGenE}] with respect to time, we find the following decoupled equations of motion for $E$-field driving:
\bea\label{eq:EoM2DEGEfield}
i\Omega u_0^{(m)}&=&-\frac{ev_{\rm{F}}}{2}\Big[\left( \delta_{m,1} + \delta_{m,-1} \right) E_x - i\left( \delta_{m,1} - \delta_{m,-1} \right) E_y \Big],\nn\\
i\Omega u_1^{(m)}+\lambda_{\rm R}u_2^{(m)}&=&0,\nn\\
i\Omega u_2^{(m)}-\lambda_{\rm R}u_1^{(m)}&=&ie\dfrac{\lambda_{\rm R}}{4k_{\rm{F}}} \Big[\left( \delta_{m,1} - \delta_{m,-1} \right) E_x-i \left( \delta_{m,1} + \delta_{m,-1} \right) E_y \Big],\nn\\
i\Omega u_3^{(m)}&=&e\dfrac{\lambda_{\rm R}}{4 k_{\rm{F}}} \Big[\left( \delta_{m,1} + \delta_{m,-1} \right) E_x - i\left( \delta_{m,1} - \delta_{m,-1} \right) E_y \Big].
\eea
Here, $E_x$ and $E_y$ are the in-plane components of $\mathbf E_0$. It is evident that the $E$-field couples only to the $m=\pm1$ harmonics. This is to be expected as $E$ excites electric dipoles, which correspond to the $|m|=1$ angular momentum channels. Equation~\eqref{eq:EoM2DEGEfield} also shows that left-circularly polarized light selectively couples to one of the $m=\pm1$ channels, while right-circularly polarized light couples to the other. We recall that $M_z$ is given in terms of the $m=0$ harmonic of $u_i$ while the in-plane magnetization are given in terms of the $m=\pm1$ harmonics of the same [cf.~Eq.~\eqref{eq:M}]. Since the $E$-field  drives only the $m=\pm 1$ harmonics, it also induces the $M_x$ and $M_y$ components of the magnetization (of amplitude $\propto \lambda_{\rm R}$) but not $M_z$.\footnote{That the $E$-field induces only the in-plane component of magnetization is valid only if the relaxation mechanism corresponds to isotropic scattering, which is what we have assumed. If scattering is anisotropic, the out-of-plane component is induced as well \cite{Engel:2007}.}

For $B$-field driving, the equations of motion read
\bea\label{eq:EoM2DEGBfield}
i\Omega u_0^{(m)}&=&0,\nn\\
i\Omega u_1^{(m)}+\lambda_{\rm R}u_2^{(m)}&=&\dfrac{g\mu_{\rm{B}}\lambda_{\rm R}}{4}\Big[\left( \delta_{m,1} + \delta_{m,-1} \right) B_x - i\left( \delta_{m,1} - \delta_{m,-1} \right) B_y \Big],\nn\\
-\lambda_{\rm R}u_1^{(m)}+i\Omega u_2^{(m)}&=&\dfrac{ g\mu_{\rm{B}}\lambda_{\rm R}}{2} \delta_{m,0} B_z,
\nn\\
i\Omega u_3^{(m)}&=&0.
\eea 

Similar to the $E$-field case, the $m=\pm 1$ harmonics are driven by the in-plane $B$-field, which translates into an induced electric current. Also like the $E$-field case, the left/right polarized $B$-field selectively couples to one of the $m=\pm 1$ harmonics. However, unlike the $E$-field case, the $B_z$ component does couple to the $m=0$ harmonic of $u$, which leads to the induced $M_z$ component of the magnetization.

Next, we follow the same steps as for the non-interacting limit to derive the equations of motion in the presence of \emph{eei}. For $E$-field driving those read
\bea\label{eq:EoM2DEG-intE}
i\Omega  u_0^{(m)}&=&-\frac{ev_{\rm{F}}}{2} \Big[(\delta_{m,1} + \delta_{m,-1}) E_x - i (\delta_{m,1} - \delta_{m,-1}) E_y \Big],\nn\\
i\Omega u_1^{(m)}+\lambda_{\rm R}u_2^{(m)}f^{(m)}_{+}+i\lambda_{\rm R}u_3^{(m)}f^{(m)}_{-}&=&0,\nn\\
i\Omega u_2^{(m)}-\lambda_{\rm R}u_1^{(m)}f^{(m)}&=&ie\dfrac{\lambda_{\rm R}}{4k_{\rm{F}}} \Big[(\delta_{m,1} - \delta_{m,-1}) E_x - i (\delta_{m,1} + \delta_{m,-1}) E_y \Big],\nn\\
i\Omega u_3^{(m)}&=&e\dfrac{\lambda_{\rm R}}{4 k_{\rm{F}}} \Big[(\delta_{m,1} + \delta_{m,-1}) E_x - i (\delta_{m,1} - \delta_{m,-1}) E_y \Big],
\eea
while for $B$-field driving we obtain
\bea\label{eq:EoM2DEG-intB}
i\Omega  u_0^{(m)}&=&0,\nn\\
i\Omega u_1^{(m)}+\lambda_{\rm R}u_2^{(m)}f^{(m)}_{+}+i\lambda_{\rm R}u_3^{(m)}f^{(m)}_{-}&=&\dfrac{g\mu_{\rm{B}}\lambda_{\rm R}}{4} \Big[(\delta_{m,1} + \delta_{m,-1}) B_x - i (\delta_{m,1} - \delta_{m,-1}) B_y \Big],\nn\\
i\Omega u_2^{(m)}-\lambda_{\rm R}u_1^{(m)}f^{(m)}&=&\dfrac{ g\mu_{\rm{B}}\lambda_{\rm R}}{2} \delta_{m,0} B_z,\nn\\
i\Omega u_3^{(m)}&=&0.
\eea
The combinations of Landau parameters in the two last equations, $f_\pm^{(m)}$ and $f^{(m)}$, are defined by Eq.~\eqref{eq:fs} of the main text. As in the main text, the symbols $\lambda_{\rm R}$, $v_{\rm F}$, and $g$ now represent renormalized quantities. In addition to these renormalizations, \emph{eei} also induces new couplings between harmonics of $u_i$, described by the last terms on the left-hand side of the second and third lines in Eqs.~\eqref{eq:EoM2DEG-intE} and \eqref{eq:EoM2DEG-intB}. However, the structure of the driving terms remain the same as in the non-interacting case. This will result in the \textit{eei} renormalization and splitting of the resonance frequencies for the in- and out-of-plane degrees of freedom. 

\section{Equations of motion for the two-valley system}\label{AppGRa}
The equations of motion for $E$-field driving are:
\bea\label{eq:EomGr1}
i\Omega u_0^{(m)}&=&-\frac{ev_{\rm{F}}}{2}\Big[(\delta_{m,1} + \delta_{m,-1}) E_x - i (\delta_{m,1} - \delta_{m,-1}) E_y \Big],\nn\\
i\Omega u_1^{(m)}+\lambda_{\rm R}u_2^{(m)}&=&0,\nn\\
-\lambda_{\rm R} u_1^{(m)}+i\Omega u_2^{(m)}+\lambda_{\rm Z}M_{3,3}^{(m)}&=&\frac{ie \lambda_{\rm R}}{4k_{\rm{F}}}\Big[(\delta_{m,1} - \delta_{m,-1}) E_x - i (\delta_{m,1} + \delta_{m,-1}) E_y \Big],\nn\\
i\Omega u_3^{(m)}-\lambda_{\rm Z}M_{3,2}^{(m)}&=&0,\nn\\
i\Omega M_{3,1}^{(m)}+\lambda_{\rm R}M_{3,2}^{(m)}&=&0,\nn\\
-\lambda_{\rm R}M_{3,1}^{(m)}+i\Omega M_{3,2}^{(m)}+\lambda_{\rm Z}u_3^{(m)}&=&0,\nn\\
i\Omega M_{3,3}^{(m)}-\lambda_{\rm Z}u_2^{(m)}&=&0.
\eea
When compared to the single valley case, even in the absence of VZ SOC, we see that there are differences in how the degrees of freedom couple to the $E$-field, e.g., the $u^{(m)}_3$ component does not couple to the $E$-field. This is a consequence of the modification of the velocity operator [Eq. \eqref{Eq:Gvel}]. Despite this, as we have seen in the main text, the current and magnetization will not be sensitive to this change. Further, with no VZ SOC, the spin-valley sector is also decoupled from the $E$-field. The presence of VZ SOC, however, couples the spin-chiral and spin-valley degrees of freedom. This suggests that in the presence of VZ SOC, one can use $E$-field to access fluctuations in the spin-valley sector.

For $B$-field driving, the equations of motion read
\bea\label{eq:EomGr2}
i\Omega  u_0^{(m)}&=&0,\nn\\
i\Omega u_1^{(m)}+\lambda_{\rm R}u_2^{(m)}&=&\dfrac{g\mu_{\rm{B}}\lambda_{\rm R}}{4} \Big[(\delta_{m,1} + \delta_{m,-1}) B_x - i (\delta_{m,1} - \delta_{m,-1}) B_y \Big],\nn\\
-\lambda_{\rm R} u_1^{(m)}+i\Omega u_2^{(m)}+\lambda_{\rm Z}M_{3,3}^{(m)}&=&\dfrac{g\mu_{\rm{B}}\lambda_{\rm R}}{2} \delta_{m,0} B_z,\nn\\
i\Omega u_3^{(m)}-\lambda_{\rm Z}M_{3,2}^{(m)}&=&0,\nn\\
i\Omega M_{3,1}^{(m)}+\lambda_{\rm R}M_{3,2}^{(m)}&=&0,\nn\\
-\lambda_{\rm R}M_{3,1}^{(m)}+i\Omega M_{3,2}^{(m)}+\lambda_{\rm Z}u_3^{(m)}&=&-\frac{ig\mu_{\rm{B}}\lambda_{\rm Z}}{4} \Big[(\delta_{m,1} - \delta_{m,-1})B_x - i (\delta_{m,1} + \delta_{m,-1}) B_y \Big],\nn\\
i\Omega M_{3,3}^{(m)}-\lambda_{\rm Z}u_2^{(m)}&=&-\frac{g\mu_{\rm{B}}\lambda_{\rm Z}}{4} \Big[(\delta_{m,1} + \delta_{m,-1}) B_x - i (\delta_{m,1} - \delta_{m,-1}) B_y \Big].
\eea
Without VZ SOC, the spin-chiral sector couples to the $B$-field in the same way as in the single-valley case, while the spin-valley sector remains decoupled. However, VZ SOC again couples these degrees of freedom. Thus, VZ SOC plays an important role in allowing both the $E$ and $B$-fields to couple to the spin-valley fluctuations.

Including \textit{eei} using Eq.~\eqref{eq:LFGr}, the equations of motion for electric-field driving are given by
\bea\label{eq:EomGr-intE}
i\Omega u_0^{(m)}&=&-\frac{ev_{\rm{F}}}{2} \Big[(\delta_{m,1} + \delta_{m,-1}) E_x - i (\delta_{m,1} - \delta_{m,-1}) E_y \Big],\nn\\
i\Omega u_1^{(m)}+\lambda_{\rm R}u_2^{(m)}f_+^{(m)}+i\lambda_{\rm R}u_3^{(m)}f^{(m)}_-&=&0,\nn\\
-\lambda_{\rm R} u_1^{(m)}f^{(m)}+i\Omega u_2^{(m)}+\lambda_{\rm Z}M_{3,3}^{(m)}h_+^{(m)}-i\lambda_{\rm Z}M_{3,2}^{(m)}h_-^{(m)}&=&ie\frac{\lambda_{\rm R}}{4k_{\rm{F}}} \Big[(\delta_{m,1} - \delta_{m,-1}) E_x - i (\delta_{m,1} + \delta_{m,-1}) E_y \Big],\nn\\
i\Omega u_3^{(m)}-\lambda_{\rm Z}M_{3,2}^{(m)}h_+^{(m)}-i\lambda_{\rm Z}M_{3,3}^{(m)}h_-^{(m)}&=&0,\nn\\
i\Omega M_{3,1}^{(m)}+\lambda_{\rm R}M_{3,2}^{(m)}h_+^{(m)}+i\lambda_{\rm R}M_{3,3}^{(m)}h_-^{(m)}&=&0,\nn\\
-\lambda_{\rm R}M_{3,1}^{(m)}h^{(m)}+i\Omega M_{3,2}^{(m)}+\lambda_{\rm Z}u_3^{(m)}f_+^{(m)}-i\lambda_{\rm Z}u_2^{(m)}f_-^{(m)}&=&0,\nn\\
i\Omega M_{3,3}^{(m)}-\lambda_{\rm Z}u_2^{(m)}f_+^{(m)}-i\lambda_{\rm Z}u_3^{(m)}f_-^{(m)}&=&0.
 \eea
Same equations for $B$-field driving  read 
\bea\label{eq:EomGr-intB}
i\Omega u_0^{(m)}&=&0,\nn\\
i\Omega u_1^{(m)}+\lambda_{\rm R}u_2^{(m)}f_+^{(m)}+i\lambda_{\rm R}u_3^{(m)}f^{(m)}_-&=&\dfrac{g\mu_{\rm{B}}\lambda_{\rm R}}{4} \Big[(\delta_{m,1} + \delta_{m,-1}) B_x - i (\delta_{m,1} - \delta_{m,-1}) B_y \Big],\nn\\
-\lambda_{\rm R} u_1^{(m)}f^{(m)}+i\Omega u_2^{(m)}+\lambda_{\rm Z}M_{3,3}^{(m)}h_+^{(m)}-i\lambda_{\rm Z}M_{3,2}^{(m)}h_-^{(m)}&=&\dfrac{g\mu_{\rm{B}}\lambda_{\rm R}}{2} \delta_{m,0} B_z,\nn\\
i\Omega u_3^{(m)}-\lambda_{\rm Z}M_{3,2}^{(m)}h_+^{(m)}-i\lambda_{\rm Z}M_{3,3}^{(m)}h_-^{(m)}&=&0,\nn\\
i\Omega M_{3,1}^{(m)}+\lambda_{\rm R}M_{3,2}^{(m)}h_+^{(m)}+i\lambda_{\rm R}M_{3,3}^{(m)}h_-^{(m)}&=&0,\nn\\
-\lambda_{\rm R}M_{3,1}^{(m)}h^{(m)}+i\Omega M_{3,2}^{(m)}+\lambda_{\rm Z}u_3^{(m)}f_+^{(m)}-i\lambda_{\rm Z}u_2^{(m)}f_-^{(m)}&=&-i\frac{g\mu_{\rm{B}}\lambda_{\rm Z}}{4} \Big[(\delta_{m,1} - \delta_{m,-1}) B_x - i (\delta_{m,1} + \delta_{m,-1}) B_y \Big],\nn\\
i\Omega M_{3,3}^{(m)}-\lambda_{\rm Z}u_2^{(m)}f_+^{(m)}-i\lambda_{\rm Z}u_3^{(m)}f_-^{(m)}&=&-\frac{g\mu_{\rm{B}}\lambda_{\rm Z}}{4} \Big[(\delta_{m,1} + \delta_{m,-1}) B_x - i (\delta_{m,1} - \delta_{m,-1}) B_y \Big].
 \eea

\bibliography{References.bib}
\end{document}